\documentclass[11pt]{elsarticle} 
\usepackage[margin=1.0in]{geometry}
\usepackage{setspace}
\usepackage{tikz}
\usepackage{graphicx}
\usepackage{pdflscape}
\usepackage{setspace}
\graphicspath{{figures/}}
\usepackage{amsmath}
\usepackage{longtable}
\usepackage{xcolor}
\usepackage{multirow}
\usepackage{amssymb}
\usepackage{tabularx}

\newcommand\clearrow{\global\let\rowmac\relax}
\clearrow

\graphicspath{{./figures/}}

\usepackage{etoolbox}
\makeatletter
\patchcmd{\ps@pprintTitle}
  {Preprint submitted}
  {Accepted}
  {}{}
\makeatother
\makeatletter
\patchcmd{\ps@pprintTitle}
  {Elsevier}
  {Icarus}
  {}{}
\makeatother

\begin{document}

\setcounter{table}{0}
\setcounter{figure}{0}

\begin{frontmatter}

\title{Emission from Volcanic SO Gas on Io at High Spectral Resolution}


\author[caltech]{Katherine de Kleer}
\author[ucb]{Imke de Pater}
\author[clemson]{M\'at\'e \'Ad\'amkovics}

\address[caltech]{Division of Geological and Planetary Sciences, California Institute of Technology, Pasadena, CA 91125}
\address[ucb]{Department of Astronomy, The University of California at Berkeley, Berkeley, CA 94720}
\address[clemson]{Department of Physics \& Astronomy, Clemson University, Clemson, SC 29634}


\end{frontmatter}

\bigbreak
\bigbreak

\section*{Abstract}
Jupiter's moon Io hosts a dynamic atmosphere that is continually stripped off and replenished through frost sublimation and volcanic outgassing. We observed an emission band at 1.707 $\mu$m thought to be produced by hot SO molecules directly ejected from a volcanic vent; the observations were made the NIRSPEC instrument on the Keck II telescope while Io was in eclipse by Jupiter on three nights in 2012-2016, and included two observations with $10\times$ higher spectral resolution than all prior observations of this band. These high-resolution spectra permit more complex and realistic modeling, and reveal a contribution to the SO emission from gas reservoirs at both high and low rotational temperatures. The scenario preferred by de Pater et al. (2002) for the source of the SO gas - direct volcanic emission of SO in the excited state - is consistent with these two temperature components if the local gas density is high enough that rotational energy can be lost collisionally before the excited electronic state spontaneously decays. Under this scenario, the required bulk atmospheric gas density and surface pressure are $n\sim 10^{11}$ cm$^{-3}$ and 1-3 nbar, consistent with observations and modeling of Io's dayside atmosphere at altitudes below 10 km (Lellouch et al. 2007; Walker et al. 2010). These densities and pressures would be too high for the nightside density if the atmospheric density drops by an order of magnitude or more at night (as predicted by sublimation-supported models), but recent results have shown a drop in SO$_2$ gas density of only a factor of 5$\pm$2 (Tsang et al. 2016). While our observations taken immediately post-ingress and pre-egress (on different dates) prefer models with only a factor of 1.5 change in gas density, a factor of 5 change is still well within uncertainties. In addition, our derived gas densities are for the total bulk atmosphere, while Tsang et al. (2016) specifically measured SO$_2$. The low-temperature gas component is warmer for observations in the first 20 minutes of eclipse (in Dec 2015) than after Io had been in shadow for 1.5 hours (in May 2016), suggesting cooling of the atmosphere during eclipse. However, individual spectra during the first $\sim$30 minutes of eclipse do not show a systematic cooling, indicating that such a cooling would have to take place on a longer timescale than the $\sim$10 minutes for cooling of the surface (Tsang et al. 2016). Excess emission is consistently observed at 1.69 $\mu$m, which cannot be matched by two-temperature gas models but can be matched by models that over-populate high rotational states. However, a detailed assessment of disequilibrium conditions will require high-resolution spectra that cover both the center of the band and the wing at 1.69 $\mu$m. Finally, a comparison of the total band strengths observed across eight dates from 1999-2016 reveals no significant dependence on thermal hot spot activity (including Loki Patera), on the time since Io has been in shadow, nor on the phase of Io's orbit at the time of observation.
%
\pagebreak


\section{Introduction}
Io is the only object in the Solar System with an atmosphere that is dynamically created by active volcanism. Interactions with Jupiter's plasma environment strip material from the atmosphere at a rate of $\sim$1 ton/second (Schneider \& Bagenal, 2007). Meanwhile, a combination of volcanic outgassing and sublimation of surface frost replenish atmospheric species, predominantly SO$_2$ (Lellouch et al. 2007). Numerous studies have investigated the role of both of these processes in sustaining Io's atmosphere, but have led to inconsistent conclusions regarding which process is dominant (e.g. Jessup and Spencer 2015; Jessup et al. 2004; Feaga et al. 2009; Spencer et al. 2005; Retherford et al. 2007; Roth et al. 2011; Tsang et al. 2012). Recent 19-$\mu$m observations have provided strong evidence that Io's SO$_2$ atmosphere decreases in column density by a factor of $\sim$5 during eclipse (Tsang et al. 2016). Although the discrepancies between various datasets remain unresolved, these authors suggest that differences could be explained if different atmospheric support mechanisms dominate at different Ionian longitudes. \par
Although Io's bulk atmosphere can be studied at wavelengths from millimeter through ultraviolet (e.g. Retherford et al. 2007; Tsang et al. 2012; Moullet et al. 2013), the direct volcanic component is challenging to isolate. In 1999, de Pater et al. (2002) detected emission near 1.7 $\mu$m while observing Io in Jupiter's shadow, and attributed the emission to the rovibronic band associated with the forbidden ${\rm a}^1\Delta \rightarrow {\rm X}^3 \Sigma^-$ transition of SO. The shape and the forbidden nature of the emission were indicative of high gas temperatures ($\geq$1000 K), and the emission was attributed to hot, excited SO gas directly released from a volcanic vent. Follow-up observations on four nights between 2000 and 2003 found a tentative correlation between the strength of the emission and activity at Loki Patera (Laver et al. 2007). De Pater et al. (2007) performed the first spatially-resolved spectroscopy of the emission band and detected a hot plume over the volcano Ra Patera, which was undergoing a powerful eruption at the time. \par
Models of the 1.7-$\mu$m emission band have treated the SO source as a single gas in thermodynamic equilibrium, with gas temperatures of 500-1000 K (de Pater et al. 2002; Laver et al. 2007; de Pater et al. 2007). However, such models were unable to fit the shape of the band, and underpredicted emission both at 1.71-1.715 $\mu$m and around 1.69 $\mu$m. The authors speculated that the source gas may be out of local thermodynamic equilibrium (LTE), but the low signal-to-noise in the wings of the band prevented more detailed modeling. \par
We observed the 1.7-$\mu$m emission band on three occasions between 2012 and 2016, during eclipses of Io by Jupiter, including two observations with a factor of $\sim$10 higher spectral resolution than all previous data. Leveraging this significant improvement to spectral resolution, and the corresponding improvement in signal-to-noise, we investigate the source of the emitting gas via more complex models than were previously warranted. The observations, data reduction and calibration procedures are described in  Section \ref{sec:data}. In Sections \ref{sec:models} and \ref{sec:results} we describe our models and results; more details on the statistical methods are given in \ref{sec:appendix}. The implications are discussed in Section \ref{sec:disc}, and summarized in Section \ref{sec:conc}.\par
\section{Observations and data reduction} \label{sec:data}
We observed Io in eclipse by Jupiter with the W. M. Keck II telescope on Mauna Kea, Hawaii, on three dates between 2012 and 2016: November 5, 2012; December 25, 2015; and May 15, 2016 (dates in UT).  On November 5 2012, we observed with the NIRSPEC instrument (McLean et al. 1998) in low-resolution mode (R$\sim$1200), and on December 25, 2015 and May 15, 2016 we observed with the same instrument in high-resolution mode (R$\sim$15000), where the structure of the line complex is significantly more resolved (although individual rotational transitions are still not resolved). During all three dates, simultaneous images of Io were obtained to identify the activity level of Io's volcanic hot spots; these data were obtained with the NIRC2 imager on Keck in 2012 and 2015 (Cantrall et al. 2018; de Kleer and de Pater 2016; de Pater et al. 2017a), and with the NIRI imager on Gemini N in 2016. In December 2015, we also obtained simultaneous data with the Keck/OSIRIS integral-field spectrograph, which spatially resolves the gas emission at low spectral resolution (de Pater et al. 2017b). In 2012 and 2015 Io was moving from sunlight into eclipse, while in 2016 Io was emerging from Jupiter occultation, and had been in shadow for an hour prior to the observation. The observing set-up varied somewhat between dates, and is described in detail for each date below. The details on the spectroscopic observations are summarized in Table \ref{tbl:obs}.
\subsection{November 5, 2012}
On UT November 5, 2012 we observed Io in eclipse with the NIRSPEC instrument. NIRSPEC is equipped with a 1024$\times$1024 InSb ALADDIN detector, and was used in low-resolution echelle spectroscopy mode for this observation, utilizing the NIRSPEC-6 spectral setting to cover the 1.62-2.02 $\mu$m region simultaneously. The peak spectral resolution in this mode is R$\sim$2500. However, by adopting the 0.76'' slit to cover as much of Io as possible, we reduce the effective spectral resolution to R$\sim$1200. Both Io and the standard star were nodded along the slit for background-subtraction. Thirteen nod pairs were obtained on Io with 30-second integration times per exposure, leading to 13 minutes of on-source time. All target exposures were used in the analysis. \par
\begin{figure}
\centering
\includegraphics[width=12cm]{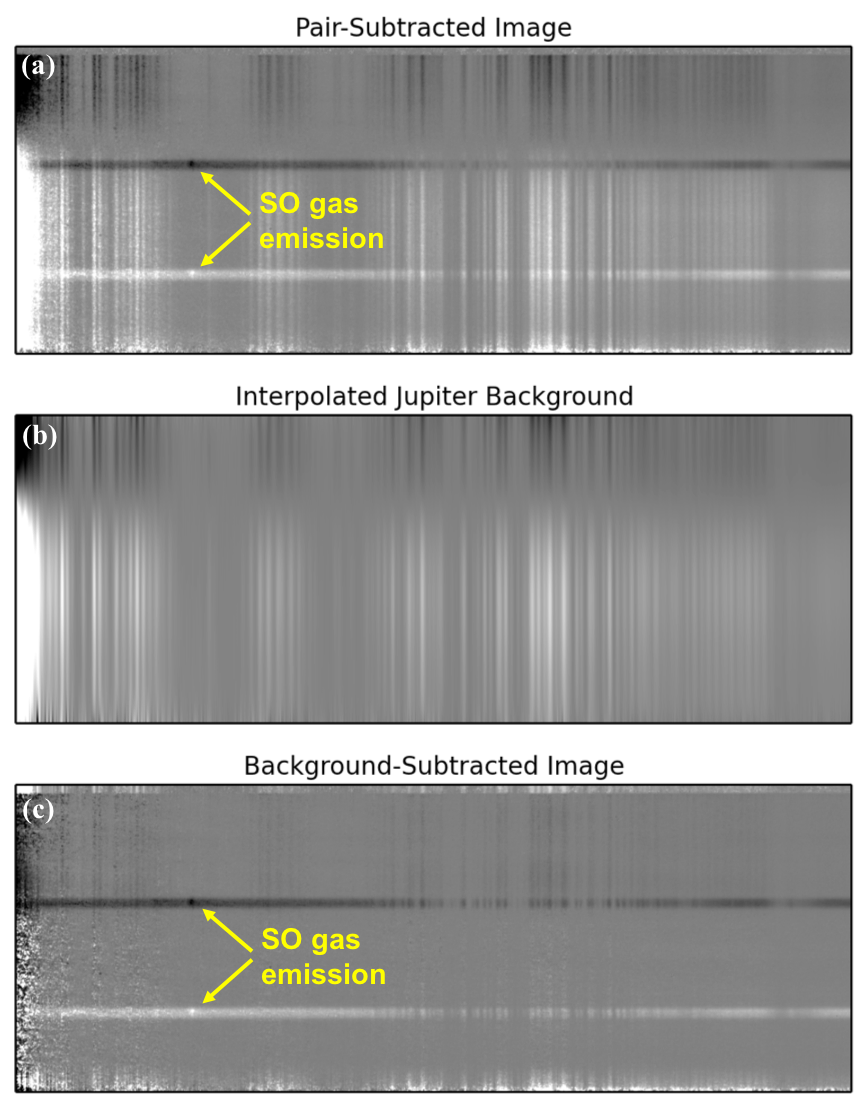}
\caption{\label{fig:bkgdsub}Demonstration of Jupiter background removal from a pair-subtracted, rectified NIRSPEC image on Nov 5, 2012. (a) Pair-subtracted image; the horizontal dark and bright lines are Io's continuum emission, and the SO line emission is indicated by arrows in both the positive and negative spectra; (b) Interpolated Jupiter background, which is not removed well in the simple pair subtraction because it varies spatially along the slit; (c) Background-subtracted image showing just the Io spectra with some residual Jupiter background. The background model is based on a polynomial fit to each column, and the scaling is the same in each frame.}
\end{figure}

The data reduction, including the spatial and spectral mapping and image rectification, was performed using the REDSPEC pipeline\footnote{UCLA infrared lab; http://www2.keck.hawaii.edu/inst/nirspec/redspec}, using telluric OH lines for spectral calibration. However, due to the uneven slit illumination caused by Jupiter, the subtraction of nodded pairs did not effectively remove background contamination, resulting in a poor pipeline extraction. Instead, the signal was extracted from a 30-pixel-wide band in the rectified images, and the 15 rows beside this band on both sides were averaged for background and subtracted from the signal. Figure \ref{fig:bkgdsub} demonstrates that the Jupiter background is smoothly varying across the slit, and can therefore be characterized well and subtracted. The stellar spectra were not subject to this contamination, and were extracted successfully via the REDSPEC software. \par
Standard stars HD 31033 (A0) and HD 32333 (G0/1V) were observed for telluric correction and photometric calibration. Model stellar spectra were created from the Castelli and Kurucz (2004) stellar atlas using the python \textbf{pysynphot} package (Lim et al. 2015). The measured stellar intensity occasionally dipped due to imperfect centering of the star within the slit and/or variable seeing conditions. Within each nod pair, if the difference in measured intensity was $>50$\% of the total measured intensity, the pair was not used in the calibration. Three out of seven stellar nod pairs were rejected based on this criterion. \par
The photometric calibration uncertainty arises from the variability in the median intensity of the stellar spectra used in the calibration (1-$\sigma$ variation of 20\%), the variability in median intensity of the Io spectra (1-$\sigma$ variation of 5.8\%), and uncertainty in the slit losses (10\%; see Section \ref{sec:slitlosses}), resulting in a total 1-$\sigma$ systematic uncertainty on the absolute flux calibration of 23\%. Within the stellar and target spectra, the pixel-to-pixel noise (residuals from a smooth curve) is at or below the 5\% level. \par
The star and Io were both at $<$1.05 airmasses throughout the observations, and the spectra were not corrected for airmass differences because the effect at 1.7 $\mu$m is negligible. The photon flux at Earth, in units of photons/s/cm$^2$/$\mu$m, was adjusted on all dates to correspond to an Io-Earth distance of 4.08 AU, as was done by Laver et al. (2007), to facilitate comparison with past work. \par
\subsection{December 25, 2015}
On UT December 25, 2015 we observed Io with both Keck telescopes simultaneously, using the NIRSPEC spectrograph in high-resolution mode (R$\sim$25,000) on Keck II (McLean et al. 1998) and the OSIRIS integral-field spectrograph with adaptive optics on Keck I (Larkin et al. 2006), for which the grating was upgraded in 2013 (Mieda et al. 2014). The latter observations were presented by de Pater et al. (2017b) and  described in detail in de Pater et al. (2018). The sky was clear during the observations but seeing was variable, in particular during the standard star observations. The observing date was selected due to the close proximity of Io to Ganymede, which was used for wavefront sensing for the adaptive optics OSIRIS observations. However, toward the end of the eclipse, Ganymede and Io were at their closest and glare from sunlit Ganymede overwhelmed Io's signal on the slit. The usable Io data on this night amounted to four nod pairs totaling twelve minutes of integration time (60-120 seconds per exposure). \par
The standard A stars HD 93702 and HD 104181 were observed for telluric correction and photometric calibration. Variable seeing during the standard star observations led to variability in the recorded count rate between exposures. Stellar nod pairs where the difference in counts recorded in the two nods was greater than 20\% were not used in the calibration. This excluded four of the twelve standard star observations. The seeing was lower and more stable during the observations of Io, and variability between nods was minimal. \par
The photometric calibration uncertainty arises from the variability in the median intensity of the stellar spectra (1-$\sigma$ variation of 20\%), the variability in median intensity of the Io spectra (1-$\sigma$ variation of $\sim$8\%), and uncertainty in the slit losses (8\%; see Section \ref{sec:slitlosses}), resulting in a total 1-$\sigma$ uncertainty on the absolute flux calibration of 23\%. The noise along the dispersion direction is under 2\% on this date, estimated from the stellar spectra. The star and Io were both at $<$1.25 airmasses throughout the observations, and the spectra were not corrected for airmass differences because the effect at 1.7 $\mu$m is at the 1\% level, well below the uncertainty from other sources. \par

\subsection{May 15, 2016}
The observations on UT May 15, 2016 utilized the NIRSPEC spectrograph in high spectral resolution mode, with a set-up identical to that of December 2015, described above. However, during the May observations thick and variable cloud cover made much of the data unusable. A high background, varying both spatially and temporally, necessitated nod pair subtraction to isolate the target spectra. We define a criterion for usable data in which the average signal-to-noise of the pair of spectra is greater than two, where signal is defined as the median level of Io's continuum above the median background, and noise is defined as the standard deviation of the sky background in the dispersion dimension, averaged over 10 spatial pixels (comparable to the extraction aperture), in the vicinity of 1.7 $\mu$m. Using this criterion, three of the seven total nod pairs were usable, amounting to twelve minutes of on-target integration time. \par
Due to the poor weather conditions, photometric calibration based on standard star data was not possible. Instead, we calibrate the data based on simultaneous Gemini N adaptive optics imaging with NIRI/ALTAIR (see Figure \ref{fig:ims_may}). The NIRI field of view included sunlit Callisto (the wavefront sensor guide star), and while Callisto was typically saturated, there were several low-transmission frames where Io was visible and Callisto was not saturated. We use these frames to calibrate the H-band intensity of Io in eclipse to that of Callisto in sunlight. We then use the spectral slope of both targets to extrapolate the 1.7-$\mu$m disk-integrated brightness of Io, assuming a 900-K Planck spectrum in the case of Io, consistent with previous spectra of Io obtained at this wavelength (Laver et al. 2007). In the frames in which Callisto was not saturated, Io was also obscured to the extent that only two of the volcanoes were visible above the noise. We thus calibrated the Io images based on these two volcanoes, and used NIRI images from an earlier part of the eclipse when cloud cover was lighter to estimate that these volcanoes contributed $\sim$2/3 of the emission seen in the less cloudy images. Taking into account these uncertainties, we find a baseline 1.7-$\mu$m intensity of Io-in-eclipse of 3-6$\times10^{-12}$ erg/s/cm$^2$/$\mu$m at Earth, and scale the NIRSPEC spectra accordingly. To facilitate comparison with past data, we scale these levels to what they would be if Io were at 4.08 AU, yielding a value of 4.6-9.2$\times10^{-12}$ erg/s/cm$^2$/$\mu$m. The range of values represent the 1-$\sigma$ uncertainty level of $\sim$33\% in absolute flux calibration. Along the spectral axis, the noise level was higher on this date than in the other observations, with a 1-$\sigma$ noise level of about 10\%. \par
\begin{table}[ht]
\begin{center}
\caption{Observations \label{tbl:obs}}
\begin{tabular}{l l l l l l l l}
\hline \hline
Date & Seeing$^a$ & Slit & Target & Time & Dist.$^b$ & Diam.$^b$ & Vel.$^c$ \\
 $[$UT$]$ & [''] & [''] & & [UT] & [AU] & [''] & km/sec \\
\hline
Nov 5, 2012 & 0.4 & 42$\times$0.760 & HD 31033 (A0) & 10:42-10:49 & & & \\
 & $>$0.4 & 42$\times$0.760 & Io sunlit & 10:52-11:29 & & \\
 & 0.4 & 42$\times$0.760 & Io eclipse & 11:37-12:10 & 4.175 & 1.208 & -10 \\
 & 0.4 & 42$\times$0.760 & HD 31033 (A0) & 12:13-12:18 & & & \\
 & 0.4 & 42$\times$0.760 & HD 32333 (G0/1V) & 12:20-12:24 & & & \\
\hline
Dec 25, 2015 & variable ($\sim$2) & 12$\times$0.720 & HD 93702 (A2V) & 13:17-13:20 & & & \\
 & 0.8-1.6 & 12$\times$0.720 & Io eclipse & 13:49-14:55 & 5.151 & 0.979 & -22 \\
 & - & 12$\times$0.720 & HD 93702 (A2V) & 15:03-15:06 & & \\
 & - &  12$\times$0.720 & HD 104181 (A0V) & 15:12-15:17 & & & \\
\hline
May 15, 2016 & \textit{0.6-1.1}$^d$ & 12$\times$0.720 & Io eclipse & 05:32-06:12 & 5.035 & 1.00 & 20 \\
 & \textit{0.6-1.1}$^d$ & 12$\times$0.720 & HD 93702 (A2V) & 06:18-06:28 & & & \\
\end{tabular}
\end{center}
\footnotesize{$^a$Seeing estimates based on data from the Mauna Kea weather center: mkwc.ifa.hawaii.edu}\\
\footnotesize{$^b$Ephemerides provided by JPL Horizons: ssd.jpl.nasa.gov/horizons.cgi}\\
\footnotesize{$^c$Io-Earth line-of-sight velocity at time of observations. Positive velocities indicate Io moving away from Earth.} \\
\footnotesize{$^d$Seeing data were not available during the May 2016 observations; the estimate is based on seeing in the hour from 07:00-08:00 UT, shortly following the observations.} \\
\end{table}
\clearpage
\subsection{Supporting imaging observations}
\begin{figure}
\centering
\includegraphics[width=18cm]{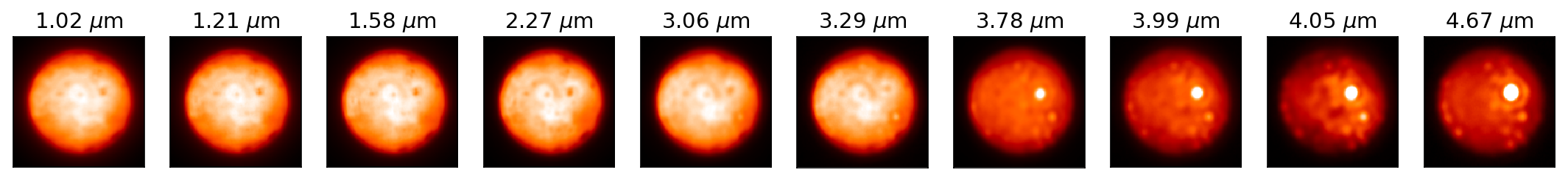}
\caption{Images of Io in sunlight on Nov 5, 2012, taken with the NIRC2 imager on Keck II just prior to the in-eclipse spectral observations. The 4.67-$\mu$m image is shown in larger size in Figure \ref{fig:ims_nirc2}, where the thermal sources are identified. \label{fig:ims_nov}}
\end{figure}

\begin{figure}
\centering
\includegraphics[width=18cm]{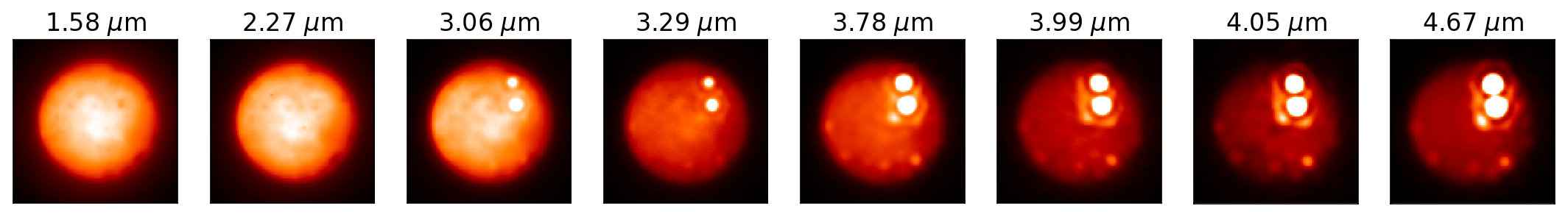}
\caption{Images of Io in sunlight on Dec 25, 2015, taken with the NIRC2 imager on Keck II just prior to the in-eclipse spectral observations. The 4.67-$\mu$m image is shown in larger size in Figure \ref{fig:ims_nirc2}, where the thermal sources are identified. \label{fig:ims_dec}}
\end{figure}

\begin{figure}
\centering
\includegraphics[width=6cm]{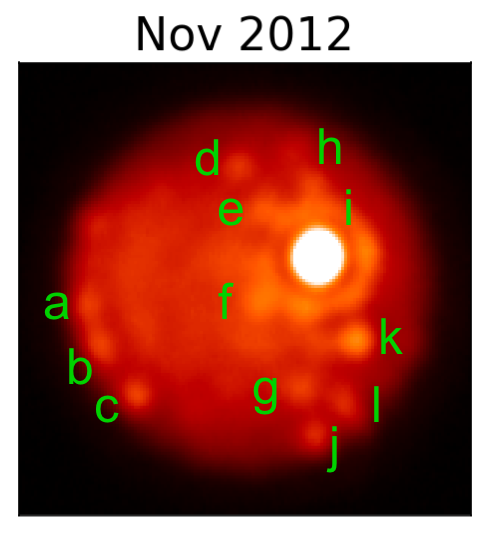}
\includegraphics[width=6cm]{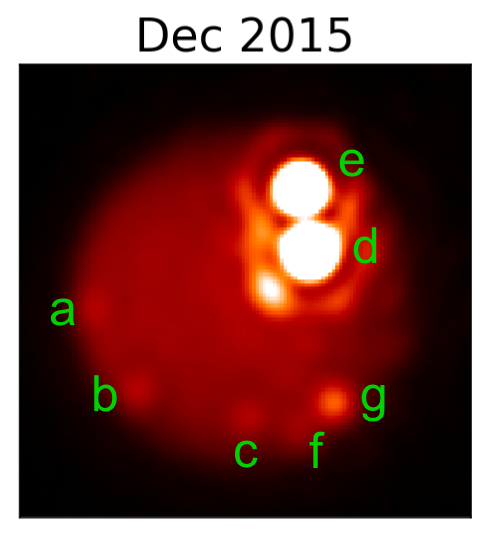}
\caption{4.67-$\mu$m images of Io on Nov 5, 2012 and Dec 25, 2015, taken with the NIRC2 imager on Keck II. Letters identifying thermal sources correspond to Tables \ref{tbl:nov_sources} and \ref{tbl:dec_sources}, which give the latitude, longitude and identification of each source on the two dates of observation. \label{fig:ims_nirc2}}
\end{figure}

\begin{figure}
\centering
\includegraphics[width=6cm]{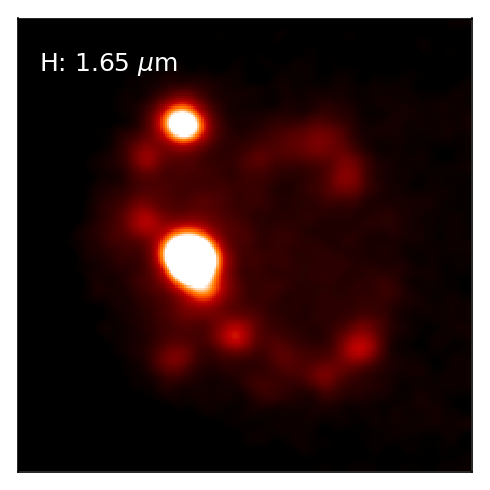}
\includegraphics[width=6cm]{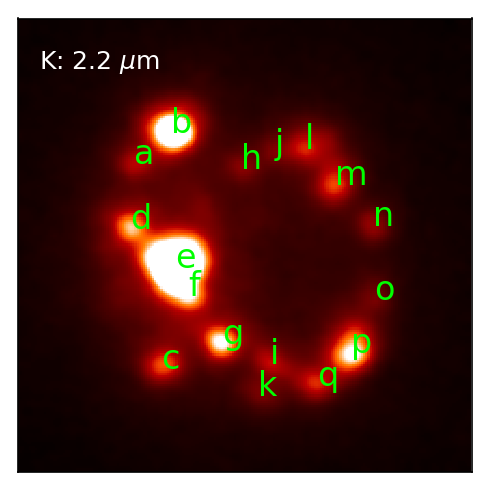}
\caption{Images of Io in eclipse on May 15, 2016 taken simultaneously with the spectra with the NIRI imager on Gemini N. The letters identifying thermal sources correspond to Table \ref{tbl:may_sources}, which gives the latitude, longitude and identification of each source. \label{fig:ims_may}}
\end{figure}

On all three dates of observation, images of Io were obtained with adaptive optics in order to identify the level of volcanic activity on the observed hemisphere. In November 2012 and December 2015 the imaging observations were made of Io in sunlight just prior to the eclipse, using the NIRC2 imager on the Keck II telescope. In May 2016, images of Io in eclipse were obtained simultaneously with the spectroscopic measurements, using the NIRI imager on the Gemini N telescope. The 2012 and 2015 hot spot intensities and identifications have been previously presented in Cantrall et al. (2018) and de Kleer \& de Pater (2016), respectively, and for Loki Patera specifically in de Pater et al. (2017a). All data were reduced and calibrated as described in de Kleer \& de Pater (2016) for each of the two instruments. The calibrated images are presented in a variety of filters in Figures \ref{fig:ims_nov}-\ref{fig:ims_may}. The hot spot labels correspond to the identifications in Tables \ref{tbl:nov_sources}-\ref{tbl:may_sources}.
%
%
%
%
%
\begin{table}
\begin{center}
\caption{Thermal Sources on November 5, 2012 \label{tbl:nov_sources}}
\begin{tabular}{c | r r | l}
\hline \hline
Source & Lat & Lon & Identification \\
 & [$^{\circ}$N] & [$^{\circ}$W] & \\
\hline
a & -3.2 & 32.2 & Janus Patera \\
b & -15.4 & 28.0 & Kanehekili Fluctus \\
c & -32.8 & 17.6 & Uta Patera \\
d & 44.5 & 335.4 & Surt \\
e & 27.9 & 327.0 & Fuchi Patera \\
f & 1.1 & 326.5 & Tol-Ava Patera \\
g & -29.5 & 311.7 & (Dark flow field) \\
h & 49.6 & 310.4 & Kinich Ahau \\
i & 12.9 & 308.1 & Loki Patera \\
j & -49.6 & 297.2 & N Lerna Regio \\
k & -13.8 & 293.2 & Gibil Patera \\
l & -35.0 & 292.2 & Ulgen Patera  \\
\hline
\end{tabular}\\
\footnotesize{From Cantrall et al. (2018)}
\end{center}
\end{table}
%
\begin{table}
\begin{center}
\caption{Thermal Sources on Dec 25, 2015 \label{tbl:dec_sources}}
\begin{tabular}{c | r r | l}
\hline \hline
Source & Lat & Lon & Identification \\
 & [$^{\circ}$N] & [$^{\circ}$W] & \\
\hline
a & -3.6 & 37.3 & Janus Patera \\
b & -34.6 & 22.7 & Uta Patera \\
c & -46.0 & 329.3 & PV170 \\
d & 15.5 & 306.4 & Loki Patera \\
e & -39.4 & 304.3 & Amaterasu Patera \\
f & -56.6 & 290.6 & N Lerna Regio \\
g & -38.8 & 285.9 & PV59 \\
\hline
\end{tabular}\\
\footnotesize{From de Kleer \& de Pater (2016)}
\end{center}
\end{table}
%
\begin{table}
\begin{center}
\caption{Thermal Sources on May 15, 2016 \label{tbl:may_sources}}
\begin{tabular}{c | r r | l}
\hline \hline
Source & Lat & Lon & Identification \\
 & [$^{\circ}$N] & [$^{\circ}$W] & \\
\hline
a & 40.4 & 87.1 & (Dark patera) \\
b & 59.4 & 83.8 & Chalybes Regio \\
c & -48.3 & 69.0 & (Dark patera) \\
d & 11.4 & 63.3 & (Dark patera) \\
e & -4.6 & 40.0 & Janus Patera \\
f & -15.5 & 35.0 & Kanehekili Fluctus \\
g & -36.1 & 22.4 & Uta Patera \\
h & 36.6 & 13.4 & (Dark patera) \\
i & -46.4 & 357.0 & Paive Patera \\
j & 44.7 & 355.0 & (Dark patera) \\
k & -68.7 & 354.9 & Inti Patera \\
l & 47.5 & 333.8 & Surt (2 sources?) \\
m & 29.5 & 326.7 & Fuchi Patera \\
n & 12.6 & 307.7 & Loki Patera \\
o & -16.2 & 303.8 & Mihr Patera \\
p & -39.6 & 299.2 & Ulgen Patera$+$PV59 \\
q & -58.9 & 294.6 & N Lerna Regio \\
\hline
\end{tabular}
\end{center}
\end{table}
\subsection{Slit losses} \label{sec:slitlosses}
Io's spatial extent exceeds the slit width, leading to significant differences in the slit losses between the calibration star and Io. The calibrated intensity of Io is therefore given by
\begin{equation}
F_{Io}=\frac{1}{f_{Io}}\times N_{Io} \times \frac{{F_*}}{\frac{1}{f_*}N_*}
\end{equation}
\begin{equation}
=\frac{f_*}{f_{Io}}\times N_{Io} \times \frac{{F_*}}{N_*},
\end{equation}
where $F$ is the flux density of the object, $N$ is the number of recorded counts/second, and $f$ is the fraction of the intensity that fell on the slit for both Io and the star ($*$). \par
\begin{figure}
\includegraphics[width=9cm]{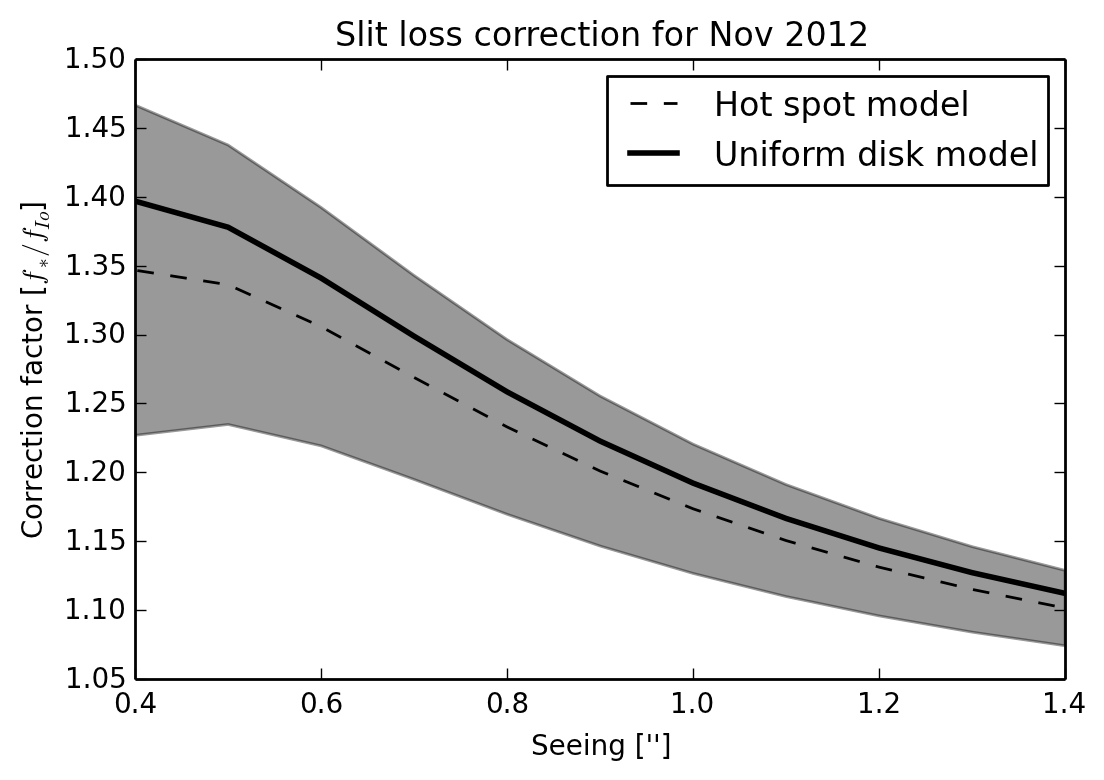}
\includegraphics[width=9cm]{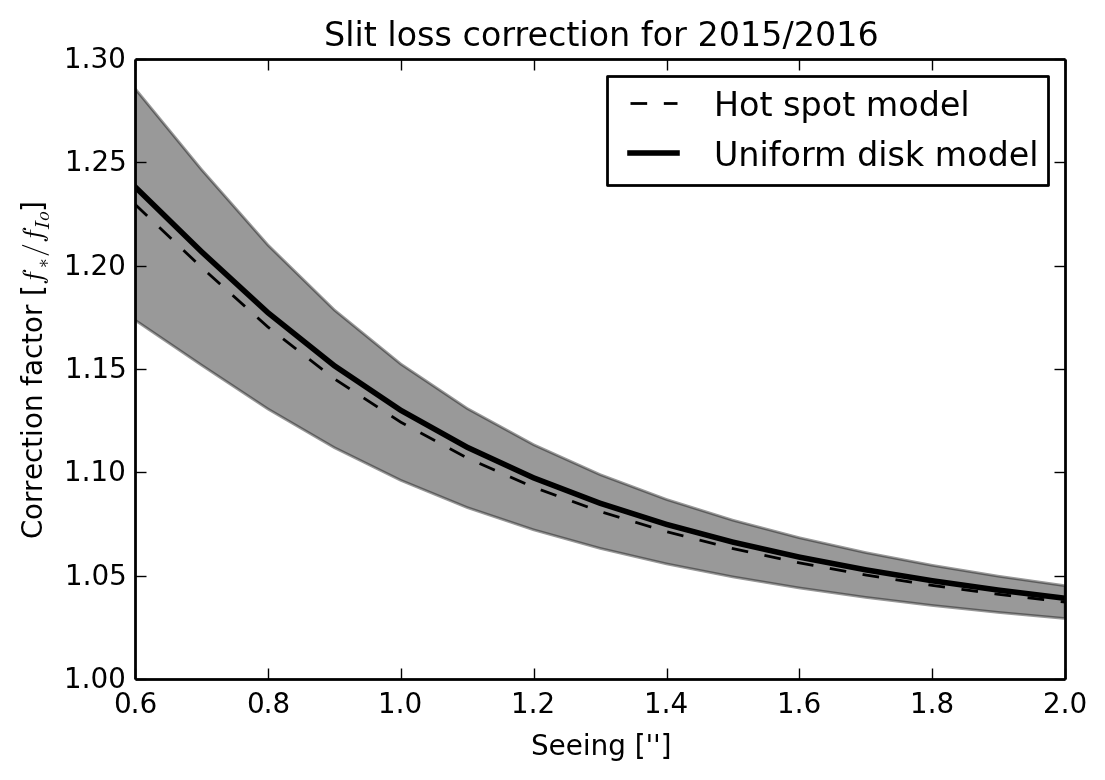}
\caption{Photometry correction factor due to difference in slit losses between the star and Io. The thick line is for an Io uniform disk model, and the shaded region, centered on the dashed line, is for models that treat Io's brightness as a random distribution of hot spots. The models differ for the two dates due to the different angular size of Io, and different slit width.\label{fig:slitlosses}}
\end{figure}

The fraction $\frac{f_*}{f_{Io}}$ is estimated by convolving models for the star and Io with a Gaussian point spread function (PSF) based on the seeing recorded on Mauna Kea during the observations (see Table \ref{tbl:obs}). The stellar model treats the star as a point source, while a range of models are used for Io to account for the unknown distribution of brightness on the disk, which may differ between the thermal and gas emission contributions. The Io models include a uniform-brightness model, as well as a range of models where 10-20 hot spots are placed randomly on the disk. The resultant distribution of slit loss correction factors are shown in Figure \ref{fig:slitlosses} as a function of seeing, for the 2012 dataset (0.76'' slit and Io angular diameter of 1.208'') and for the 2015 and 2016 datasets (0.72'' slit and Io angular diameter of 0.979-1.00''). The shaded region on each plot indicates the uncertainty in the correction factor $\frac{f_*}{f_{Io}}$ based on the uncertain brightness distribution. These uncertainties are based on the assumption that the slit was aligned along the center of Io's disk; they may be underestimated if Io was not centered on the slit, as could happen if the thermal emission was asymmetric across the disk and biased guiding. Based on the seeing on Mauna Kea during the time of observations, given in Table \ref{tbl:obs}, the slit loss correction factors are in the range of 1.2-1.45 for the 2012 observations, 1.05-1.2 for the 2015 observations, and 1.1-1.3 for the 2016 observations. \par
%
\subsection{Doppler correction}
For the low-resolution data from 2012, the line-of-sight velocity was around -10 km/s (Io moving toward Earth), and changed by $\sim$1.5 km/s during the event. The average Doppler correction is therefore around 5$\times 10^{-5}$ $\mu$m, much smaller than the wavelength sampling of 4$\times 10^{-4}$ $\mu$m. During the high-resolution observations, the line of sight velocity was around 20 km/s (Io moving toward Earth in 2016 and away from Earth in 2015). The Doppler shift of 1$\times 10^{-4}$ $\mu$m is several spectral elements (of size 2.35$\times 10^{-5}$ $\mu$m). The $<$1 km/s change in the line-of-sight velocity during each observation is still well below the spectral resolution. Nevertheless, on each date, the spectra of Io were each individually tellurically-corrected and Doppler-corrected before being median-combined. \par
\section{Analysis} \label{sec:models}
\subsection{Thermal continuum}
\begin{figure}
\centering
\includegraphics[width=16cm]{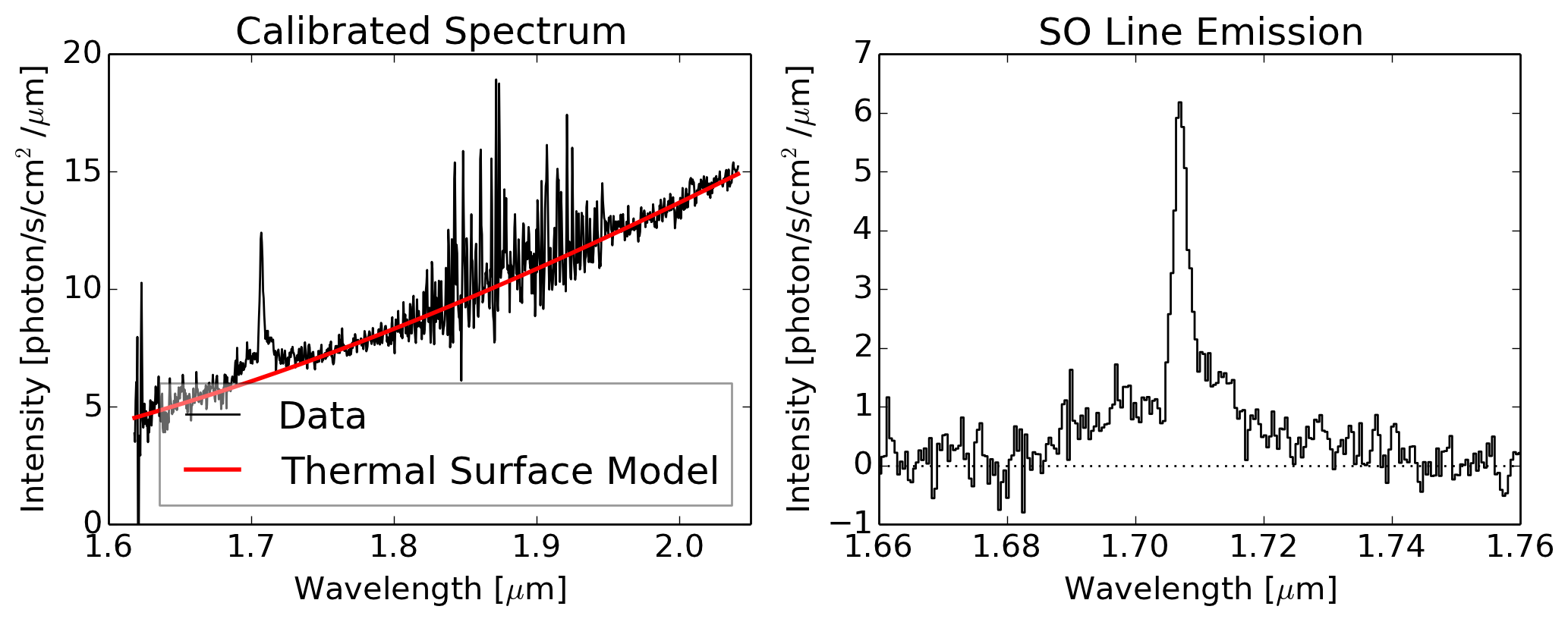}
\caption{Low-resolution Io spectrum from November 5, 2012. (Left) Full observed spectrum with best-fit model for thermal continuum; and (Right) Spectrum covering only the SO emission band, with thermal continuum subtracted. \label{fig:spec_2012}}
\end{figure}

In the near-IR, Io's in-eclipse spectrum is dominated by thermal emission from active volcanism. In the low-resolution dataset, this background is estimated by fitting the 1.6-2.0 $\mu$m spectrum excluding the SO emission region (see Figure \ref{fig:spec_2012}). For the high-resolution datasets, the spectral coverage lacks sufficient baseline regions, and the thermal emission is left as a free parameter in the fits to the gas emission band. The thermal emission is modeled with a Planck function at a single temperature and emitting area ($T_{th},A_{th}$); this simple parameterization provides a good fit to the data, and the resultant temperatures fall within the range seen in previous observations with larger spectral coverage ($\sim$500-1000 K; de Pater et al. 2002; Laver et al. 2007), as well as with the temperature estimated for December 25, 2015 from the OSIRIS dataset (de Pater et al. 2018). 
\subsection{Gas emission models}
The SO emission band near 1.7 $\mu$m was first identified on Io by de Pater et al. (2002), and is produced by the 0-0 vibrational band of the forbidden electronic transition from the metastable ${\rm a}^1\Delta$ excited state to the ${\rm X}^3\Sigma^-$ ground state. Setzer et al. (1999) identified 225 individual rovibronic lines over nine rotational branches within this band. \par
The SO emission line models are calculated for a given gas temperature according to the method described in de Pater et al. (2002) and Laver et al. (2007), and convolved with a Gaussian PSF to match the spectral resolution of the instrument. For the high-resolution datasets, the spectral resolving power used for the model convolution is:
\begin{equation}
R_{0.72''slit} = R_{0.43''slit}\times \frac{0.43''}{0.72''}.
\end{equation}
\indent $R_{0.43''slit}=$25,000, leading to a spectral resolution of R$\sim$15,000 for the 0.72'' slit. The low-resolution observations have a spectral resolution of R$\sim$1250 after degrading from the nominal 0.38'' slit (where R$\sim$2500) to the 0.76'' slit.\par
Three gas emission models are fit to the data:
\begin{enumerate}
\item[]{\textbf{1T}: a single gas temperature, with the entire source in local thermodynamic equilibrium.}
\item[]{\textbf{2T}: two gas sources at different temperatures, each in local thermodynamic equilibrium.}
\item[]{\textbf{nLTE(A)}: high and low rotational states populated according to different temperatures.}
\item[]{\textbf{nLTE(B)}: background 2T atmosphere with excess population in high rotational states.}
\end{enumerate}\par
A summary of these models and their free parameters is given in Table \ref{tbl:fits}.\par
\begin{table}
\footnotesize
\begin{center}
\caption{Model parameters \label{tbl:fits}}
\begin{tabular}{l l l l}
\hline 
Model & Scenario & Free Parameters$^a$ & Model \\
\hline
1T & One source in thermal equilibrium & \{$T$,$c$,$A_{th}$,$T_{th}$\} & $cF(T)+A_{th}B_{\lambda}(\lambda,T_{th})$ \\
2T & Two sources, each in thermal equilibrium & \{$T_1$,$T_2$,$c_1$,$c_2$,$A_{th}$,$T_{th}$\} & $c_1F(T_1)+c_2F(T_2)+A_{th}B_{\lambda}(\lambda,T_{th})$ \\
nLTE(A) & Thermal disequilibrium (Model A) & \{$T_1$,$T_2$,$c_1$,$c_2$,$J$\} & $c_1F(T_1,J)+c_2F(T_2,J)$ \\
nLTE(B) & Thermal disequilibrium (Model B) & \{$T_3$,$c_1$,$c_2$,$c_3$,$J$\} & $c_1F(T_1)+c_2F(T_2)+c_3F(T_3,J)$\\
\hline
\end{tabular}
\end{center}
\footnotesize{$^a$In the low-resolution data from November 2012, the background thermal emission was well constrained due to the larger spectral coverage, and the parameters $A_{th}$ and $T_{th}$ were determined and surface emission subtracted prior to the gas model fits.}
\end{table}
\textbf{One-temperature (1T) models:} These models represent the simplest family of gas emission models, and treat the emission as arising from gas in local thermodynamic equilibrium at a single temperature. The free parameters are the gas temperature $T$ and a scale factor $c$, in addition to the emitting area and temperature of the thermal continuum ($A_{th}$ and $T_{th}$). For all models, scale factors are used to match the models to the absolute photon contribution from each temperature component; the quoted results give integrated band strength instead of the value of the scale factors, as this is the more physically meaningful value. On all dates, the 1T models deviated from the data systematically near the band peak, leading to the more complex models described below. \par
\textbf{Two-temperature (2T) models:} The two-temperature models treat the SO gas as two reservoirs of molecules, each in local thermodynamic equilibrium at a different temperature. Preliminary fits to all datasets using a grid of temperatures demonstrated a consistent preference for a distinct low-temperature ($\lesssim$few hundred K) and high-temperature ($>$500 K) component. In order to avoid parameter degeneracy, we therefore restrict the potential range of one temperature component to $<$500 K, and the other to $>$500 K.\par
\textbf{Disequilibrium (nLTE) models:} The 2T models provide an excellent fit to the high-resolution data, but are not able to match the shape of the emission band farther from the line center, in the spectral region covered by the low-resolution dataset from 2012 but not the high-resolution datasets from 2015 and 2016. In order to fit the full band shape in the 2012 data, we postulate a gas out of local thermodynamic equilibrium (non-LTE). In particular, we allow an over-population of molecules at high rotational quantum numbers ($J$). A thorough exploration of non-LTE models would require more free parameters than can be uniquely fit by the low-resolution dataset, so instead we treat the non-LTE scenario with two simple parameterizations:\par
\textit{(Model A) Distinct temperatures for high and low rotational states:} In this treatment, we model the emission as arising from a gas population where the high and low rotational levels are populated according to Boltzmann distributions at different temperatures. We test a grid of models as follows: the inputs to each model are the temperatures according to which the low-$J$ and high-$J$ states are populated, the $J$-value cut-off between the two temperature populations, and the relative population at each temperature. We test a grid of temperatures and $J$-cutoff, fixing the value of each of these parameters at each point in the grid and fitting only for the relative populations. The temperatures are tested at 50 K increments between 50 and 2000 K, and all $J$-cutoff values from 10 through 59 are tested. \par
\textit{(Model B) Fixed background atmosphere with added population in high-J states:} In this treatment, we fix the background atmosphere following the 2T models that find the best fit to the high-resolution data: the two component gases are fixed at temperatures of 170 K and 1000 K. The former temperature is consistent with the temperature found for the low-T component in Dec 2015, when Io was also moving from sunlight into eclipse, as well as with previous measurements of Io's atmospheric temperature in sunlight (Lellouch et al. 2015). On top of this, we add a population of molecules at high rotational states only, populated according to a specific temperature that is independent of the background atmosphere temperatures. The inputs to the model are scaling factors for the low-T, high-T, and non-LTE components, the temperature according to which the high-$J$ states are populated, and the $J$ value above which over-population occurs. As in the previous case, we test a grid of temperatures and $J$-cutoff, fixing the value of each of these parameters at each point in the grid and fitting only for the relative populations. The temperatures are tested at 50 K increments between 50 and 2000 K, and all $J$-cutoff values from 20 through 59 are tested. \par
We note that these models are only two of many possible ways to parameterize a non-LTE state, and that the grid method does not allow for a robust statistical assessment of uncertainties on best-fit parameters. The best-fit parameter values should therefore be viewed with some caution, but the resultant models demonstrate that removing the LTE restriction allows for models that are capable of matching the broad-scale shape of the entire emission band. \par
\subsection{Model fitting and selection criteria}
In all cases where fits are performed, the best-fit parameter values are determined through a likelihood optimization routine, while the uncertainties are estimated using an affine-invariant ensemble Markov Chain Monte Carlo (MCMC) simulation. These methods are described in detail in \ref{sec:appendix}, where the methodology is shown for the specific example of 2T model fits to the high-resolution December 2015 data. \par
%
We calculate two metrics for the purposes of assessing the quality of the fit and comparing between models: the $\chi^2$ value, and the autocorrelation of the residuals. The $\chi^2$ value is normalized to $N_{data}$-$N_{par}$, where $N_{data}$ is the number of data points and $N_{par}$ is the number of free parameters in a given fit. We present the value of this metric alongside our final models to provide a quantitative evaluation of the goodness-of-fit. However, we note that for non-linear models the number of degrees of freedom does not correspond directly to $N_{data}$-$N_{par}$, and the interpretation of this metric for the purposes of model comparison is therefore not straightforward (Andrae et al. 2010). \par
The second metric we calculate for each model is the autocorrelation of the residuals. Under the assumption that the noise in the data is random and not correlated in wavelength, the residuals should also exhibit no correlation between different wavelengths in the case of a perfect model fit, while an incorrect model will lead to systematic deviations from the data and hence a positive autocorrelation of the residuals. We calculate the autocorrelation between datapoints separated in wavelengths using the Pearson R coefficient, which is the ratio of the covariance between the two datasets and the product of the standard deviations of each dataset:
\begin{equation}
R_{a,b} = \frac{cov(a,b)}{\sigma_a \sigma_b}.
\end{equation}
Here, the two datasets are the residuals from a given model fit, and the same residuals offset by $n$ pixels; we calculate this coefficient for offsets from one to five pixels. \par
\section{Results} \label{sec:results}
\begin{figure}
\includegraphics[width=10cm]{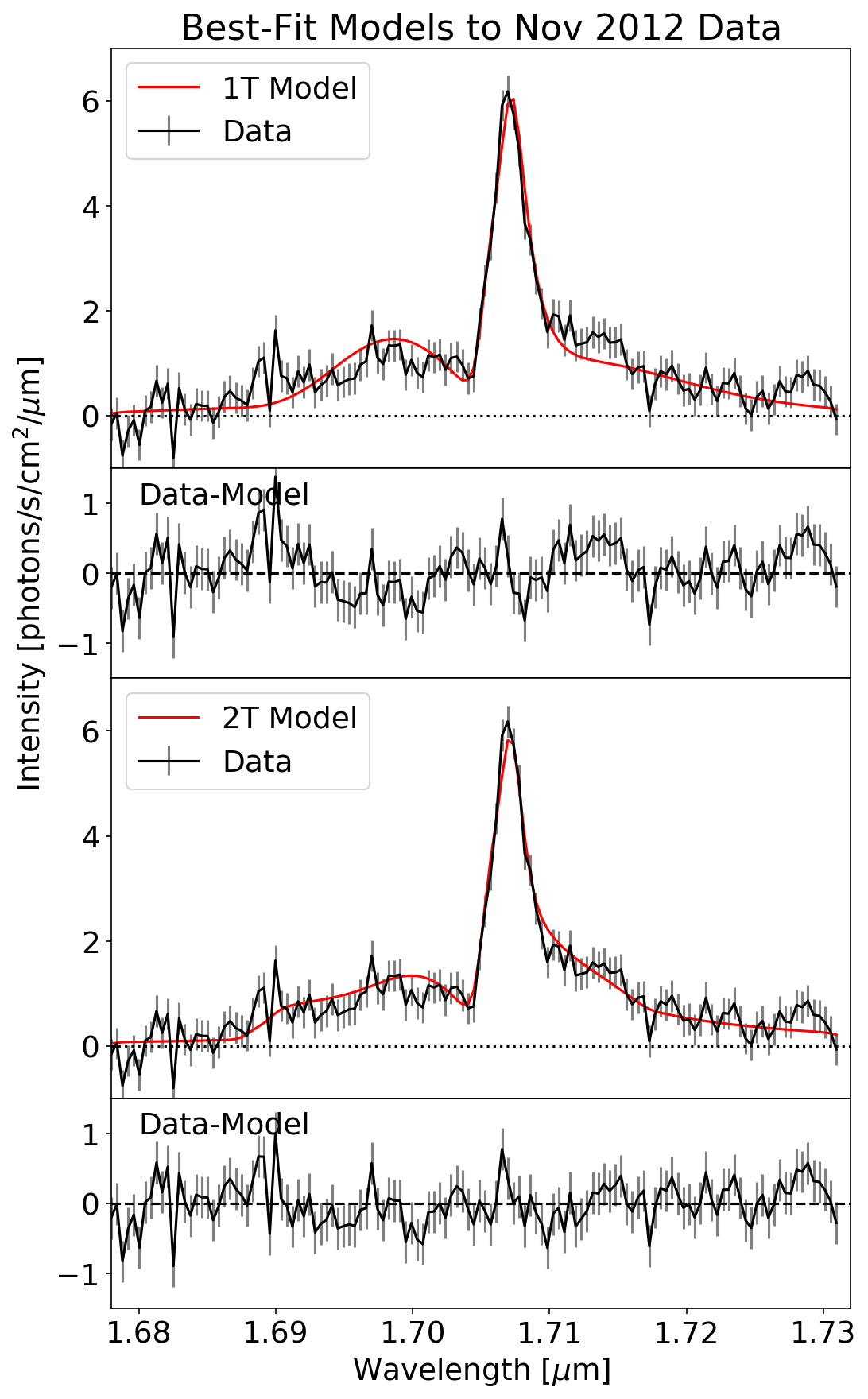}
\caption{Best-fit one- and two-temperature models to the November 5, 2012 dataset, with residuals. Although the 2T model provides a better fit to the band shape, neither model can reproduce all of the features of the observations, in particular the excess of emission around 1.69 and 1.73 $\mu$m. The parameters corresponding to the models shown here are given in Table \ref{tbl:results_fits}, and goodness-of-fit metrics are tabulated in Table \ref{tbl:stats}. \label{fig:spec_lowres}}
\end{figure}

\begin{figure}
\includegraphics[width=18cm]{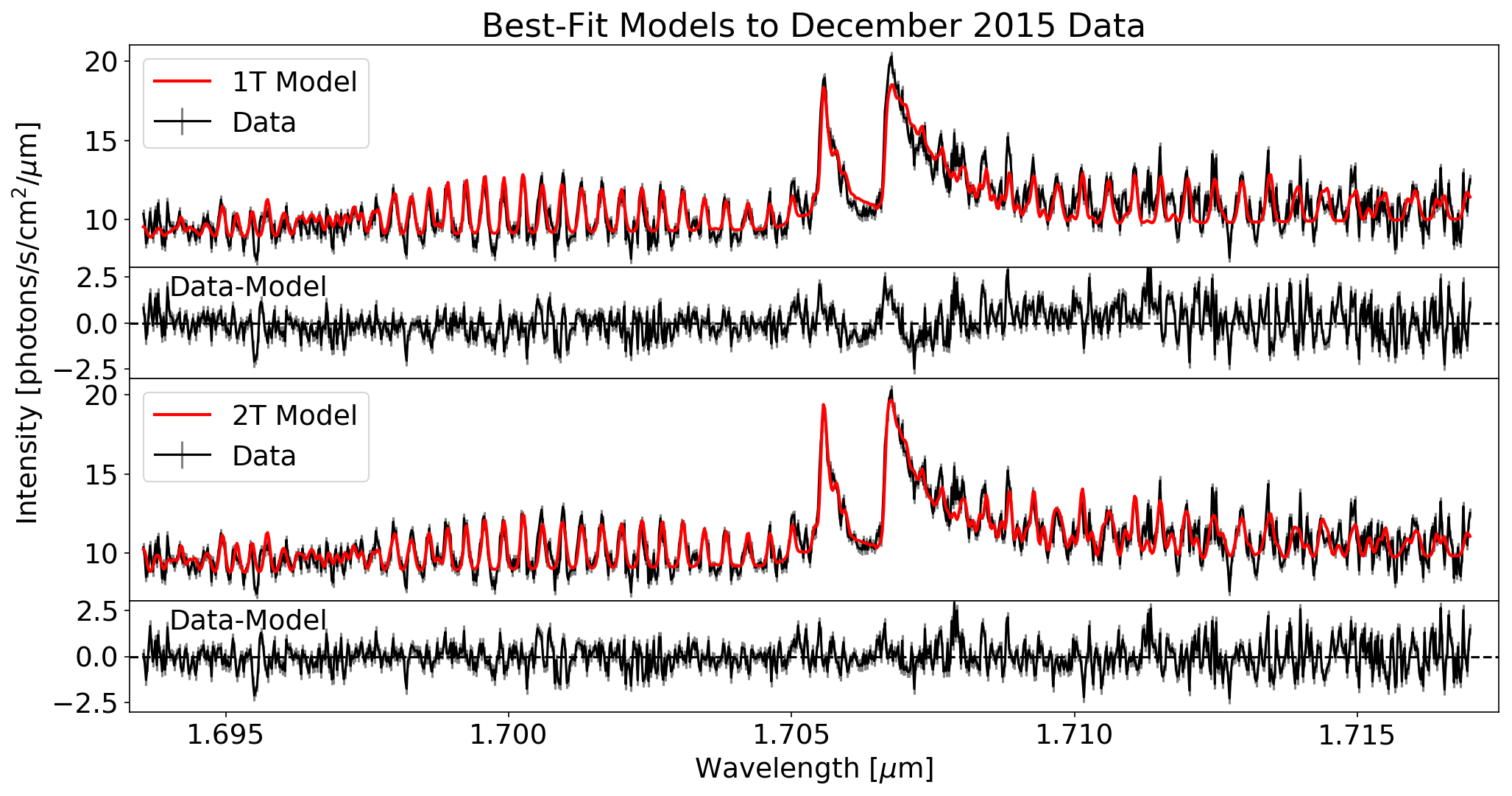}
\includegraphics[width=18cm]{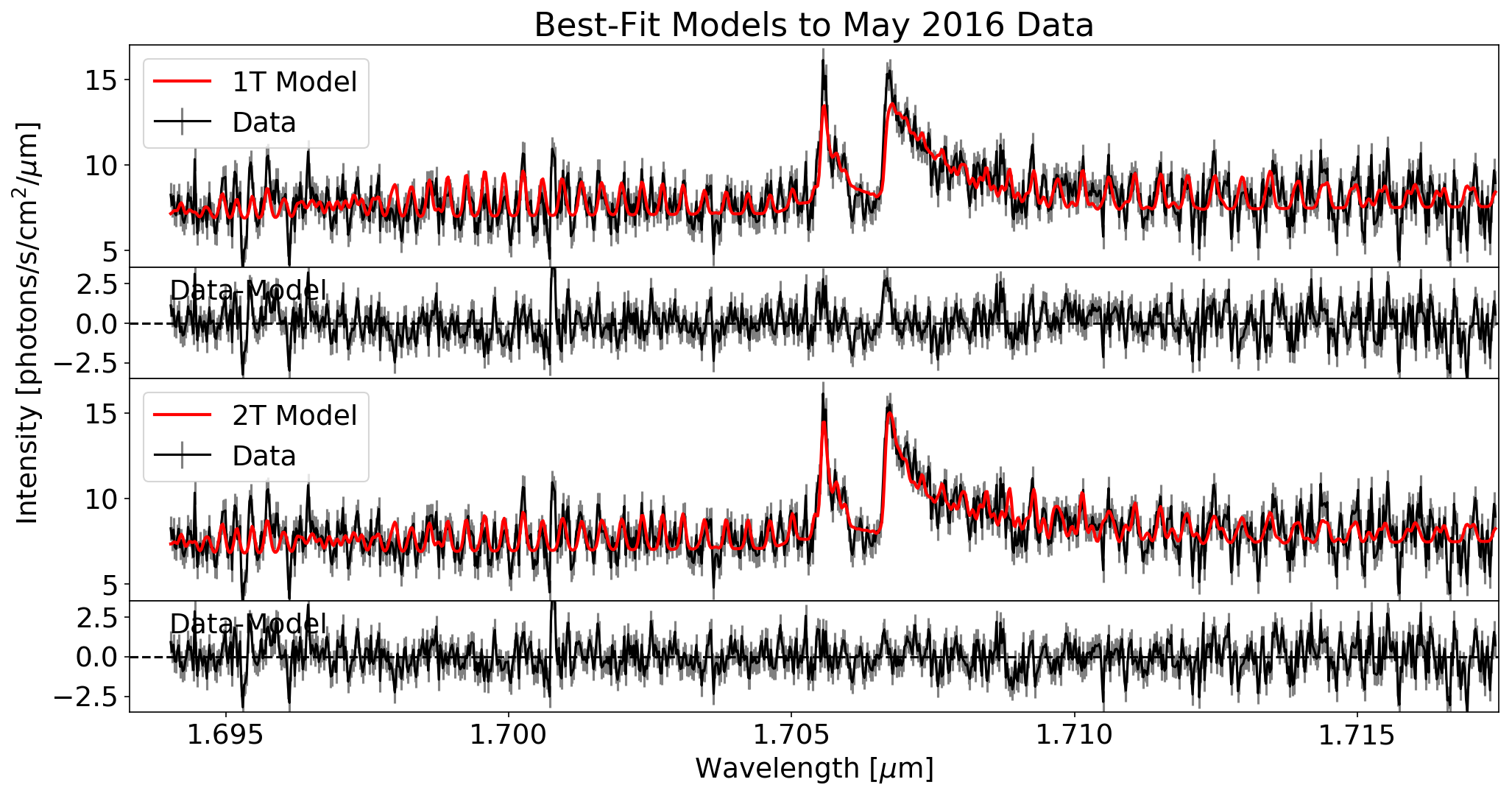}
\caption{Best-fit 1T and 2T models to high-resolution data from December 25, 2015 and May 15, 2016, with residuals. On both dates, the 1T models show deviations from the data near the band peaks, which are better matched by the 2T models. The parameters corresponding to the models shown here are given in Table \ref{tbl:results_fits}, and goodness-of-fit metrics are given in Table \ref{tbl:stats}. \label{fig:spec_highres}}
\end{figure}

The best-fit one- and two-temperature gas models for all three datasets are shown in Figures \ref{fig:spec_lowres} and \ref{fig:spec_highres}, and the corresponding best-fit thermal and gas temperatures are given in Table \ref{tbl:results_fits}. The normalized $\chi^2$ value and R coefficient for autocorrelation of the residuals are reported in Table \ref{tbl:stats}. The latter is reported for wavelength intervals of $n$=1-5 pixels; R values not consistent with zero indicate that the residuals are correlated across $n$ datapoints, an indication of systematic deviations between model and data rather than random noise. For all three dates, the 1T models provide poorer fits, as seen in the higher $\chi^2$ and in the systematic deviations from the data near the band peak, which can be seen qualitatively in the figures and quantitatively through the values of the Pearson R coefficient, which indicates a greater correlation between wavelengths for the 1T models than for the other models tested. \par
\begin{figure}
\includegraphics[width=18cm]{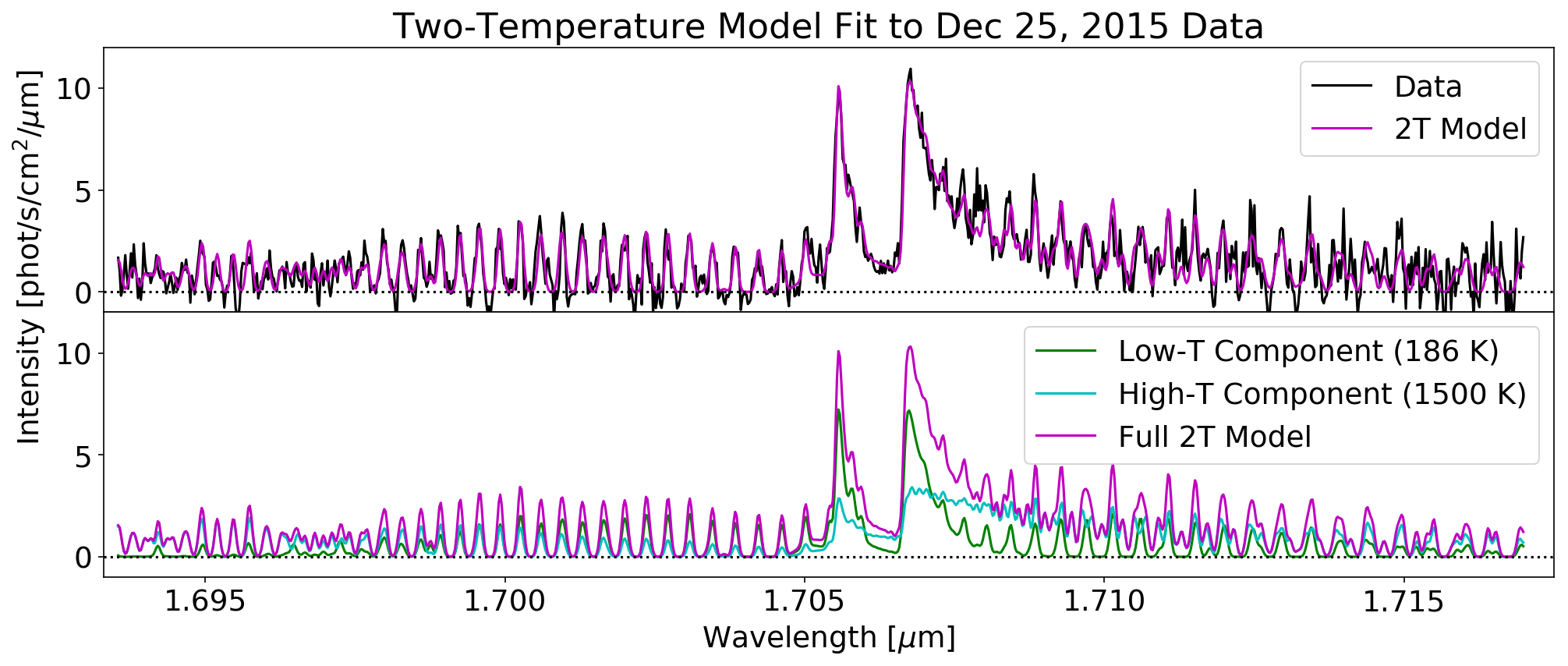}
\caption{Observed spectrum from Dec 25, 2015 with 2T model. The bottom plot shows the breakdown of the model into the two temperature components, demonstrating qualitatively which features of the spectrum are produced by each temperature component. \label{fig:2Tdemo}}
\end{figure}

The 2T models provide an excellent fit to the high-resolution data from December 2015, though the improvement over the 1T models is less significant for the May 2016 data. On both dates, the best-fit temperature values include a low-temperature component around 100-200 K and a high-temperature component of $\gtrsim$1000 K. The need for both a low and high temperature component to match the observed band shape can be seen qualitatively in Figure \ref{fig:2Tdemo}, where the contribution from each component is shown; the cool component produces the narrow band peaks, while the hot component produces the shallower shoulders on the main peaks and is needed to produce the emission seen around 1.695 $\mu$m. \par
%
%
%
%
\begin{table}
\begin{center}
\caption{Model Parameters from Fits \label{tbl:results_fits}}
\begin{tabular}{>{\rowmac}l|>{\rowmac}l|>{\rowmac}l>{\rowmac}l<{\clearrow}}

\hline \hline
\multicolumn{4}{c}{One-Temperature Models}\\
\hline
Dataset & Parameter & Maximum Likelihood & Credible Interval \\
\hline
Nov 2012 & $T_{Th}$ [K]$^a$ & 870 & (864, 876) \\
 & $T_{SO}$ [K] & 591 & (503, 746) \\
 \hline
Dec 2015 & $T_{th}$ [K] & 631 & (627, 645) \\
 & $T_{SO}$ [K] & 465 & (441, 488) \\
 \hline
May 2016 & $T_{th}$ [K] & 742 & (706, 790) \\
 & $T_{SO}$ [K] & 494 & (465, 531) \\
 \hline
 \multicolumn{4}{c}{Two-Temperature Models}\\
\hline
Nov 2012 & $T_{Th}$ [K]$^a$ & 870 & (864, 876) \\
 & $T_{SO,1}$ [K] & 283 & (214, 422)  \\
 & $T_{SO,2}$ [K] & 1605 & (940, 1804) \\
 \hline
Dec 2015 & $T_{th}$ [K]  & 644 & (642, 684) \\
 & $T_{SO,1}$ [K]  & 186 & (172, 217) \\
 & $T_{SO,2}$ [K]  & 1500 & (1395, 1807) \\
 \hline
 May 2016 & $T_{th}$ [K] & 727 & (681, 729)  \\
 & $T_{SO,1}$ [K] & 91 & (76, 119) \\
 & $T_{SO,2}$ [K] & 971 & (866, 1105) \\
\hline
\end{tabular}
\end{center}
\footnotesize{$^a$For the Nov 2012 observations, $T_{Th}$ was determined from regions of the spectrum free from gas emission, and the thermal surface spectrum was subtracted from the data prior to fitting the gas model. On the other dates, the surface and gas temperatures were fit for simultaneously.}
\end{table}
%
\begin{table}
\footnotesize
\begin{center}
\caption{Goodness-of-fit metrics \label{tbl:stats}}
\begin{tabular}{l | l | l | lllllll}
\hline
Date & Model & $\chi^2_N$ & \multicolumn{5}{c}{Pearson R statistic for autocorrelation of residuals, with 95\% confidence intervals} & \\
 & & & \multicolumn{1}{c}{$\Delta \lambda=1$} & \multicolumn{1}{c}{2} & \multicolumn{1}{c}{3} & \multicolumn{1}{c}{4} & \multicolumn{1}{c}{5} & \\
\hline
Nov 2012 & 1T & 1.54 & 0.41 [0.26,0.55] & 0.37 [0.21,0.51] & 0.28 [0.12,0.43] & 0.28 [0.11,0.43] & \textit{0.13 [-0.03,0.30]} & \\
 & 2T & 1.15 & 0.24 [0.07,0.39] & 0.21 [0.04,0.37] & \textit{0.12 [-0.04,0.29]} & \textit{0.14 [-0.03,0.30]} & \textit{-0.01 [-0.19,0.15]} & \\
  & nLTE(A) & 1.05 & 0.18 [0.01,0.34] & \textit{0.16 [0.00,0.33]} & \textit{0.09 [-0.08,0.25]} & \textit{0.14 [-0.03,0.30]} & \textit{0.00 [-0.17,0.17]} & \\
 & nLTE(B) & 1.05 & 0.18 [0.01,0.34] & \textit{0.17 [0.00,0.33]} & \textit{0.08 [-0.09,0.25]} & \textit{0.10 [-0.07,0.27]} & \textit{-0.05 [-0.22,0.11]} & \\
 \hline
Dec 2015 & 1T & 1.26 & 0.47 [0.42,0.52] & 0.22 [0.16,0.28] & 0.10 [0.03,0.16] & \textit{0.06 [0.00,0.12]} & 0.11 [0.05,0.17] & \\
 & 2T & 1.01 & 0.35 [0.29,0.40] & \textit{0.06 [0.00,0.12]} & \textit{-0.06 [-0.12,0.00]} & \textit{-0.06 [-0.12,0.00]} & \textit{0.01 [-0.04,0.07]} & \\
 \hline
May 2016 & 1T & 2.18 & 0.49 [0.44,0.54] & 0.19 [0.13,0.25] & 0.08 [0.02,0.14] & \textit{0.01 [-0.05,0.07]} & \textit{-0.06 [-0.12,0.00]} & \\
 & 2T & 2.01 & 0.45 [0.40,0.50] & 0.13 [0.07,0.19] & \textit{0.03 [-0.02,0.09]} & \textit{-0.02 [-0.08,0.03]} & -0.08 [-0.15,-0.02] & \\
\hline
\end{tabular}
\end{center}
Notes: $\chi^2_N=\chi^2/(N_{data}-N_{par})$, the value of $\chi^2$ normalized by the number of datapoints minus the number of free parameters in the fit. The Pearson R statistic is consistent with zero if there is no significant autocorrelation in the residuals. The low and high confidence intervals are given in brackets, and italics indicate lags with no significant autocorrelation. \\
\end{table}
%
The low-resolution data from November 2012 have larger spectral coverage, and show an excess of emission around 1.69 $\mu$m also seen in previous datasets (de Pater et al. 2002; Laver et al. 2007) that cannot be captured by the equilibrium models. Two disequilibrium model families are tested, one in which the entire emission band arises from a gas population where the high and low rotational states are populated according to independent temperatures, and one that places an excess population at high rotational states on top of a two-temperature model with fixed high and low temperatures of 1000 and 170 K. Both models provide equally good fits to the data, and improve on the standard 2T models for the November 2012 data (see Table \ref{tbl:stats}). In particular, disequilibrium models that over-populate molecules with rotational quantum numbers above $J\sim 35$ do a better job of capturing the overall shape of the band, including the excess of emission around 1.69 $\mu$m. In addition, some of these models produce an excess of emission around 1.725-1.73 $\mu$m where excess emission at the few-$\sigma$ level is also seen in the data. These models demonstrate the ability of disequilibrium gas states to resolve the discrepancy between the simpler 2T models and the data. Future data with higher spectral resolution across this entire band would allow for a more in-depth modeling and fitting procedure. \par
\begin{figure}
\centering
\includegraphics[width=10cm]{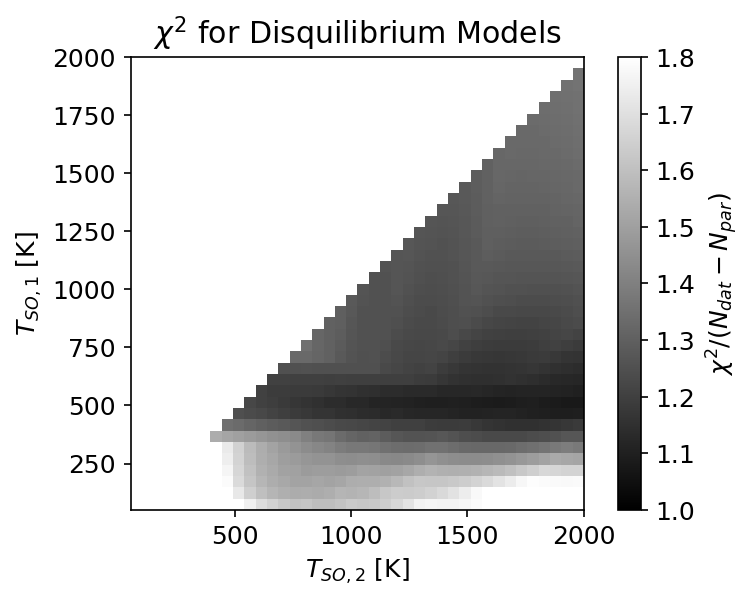}
\caption{Chi-squared values for best-fit models to the November 2012 data using a grid of disequilibrium models (nLTE Model A), normalized by the number of datapoints minus the number of free parameters in the fit. The lowest values of $\chi^2$ are obtained for $J$ levels below $\sim$35 populated according to a temperature around 500 K, with higher rotational states overpopulated. The temperature of the high-T component is poorly constrained by the data. \label{fig:x2grid_nLTE}}
\end{figure}
\begin{figure}
\centering
\includegraphics[width=8cm]{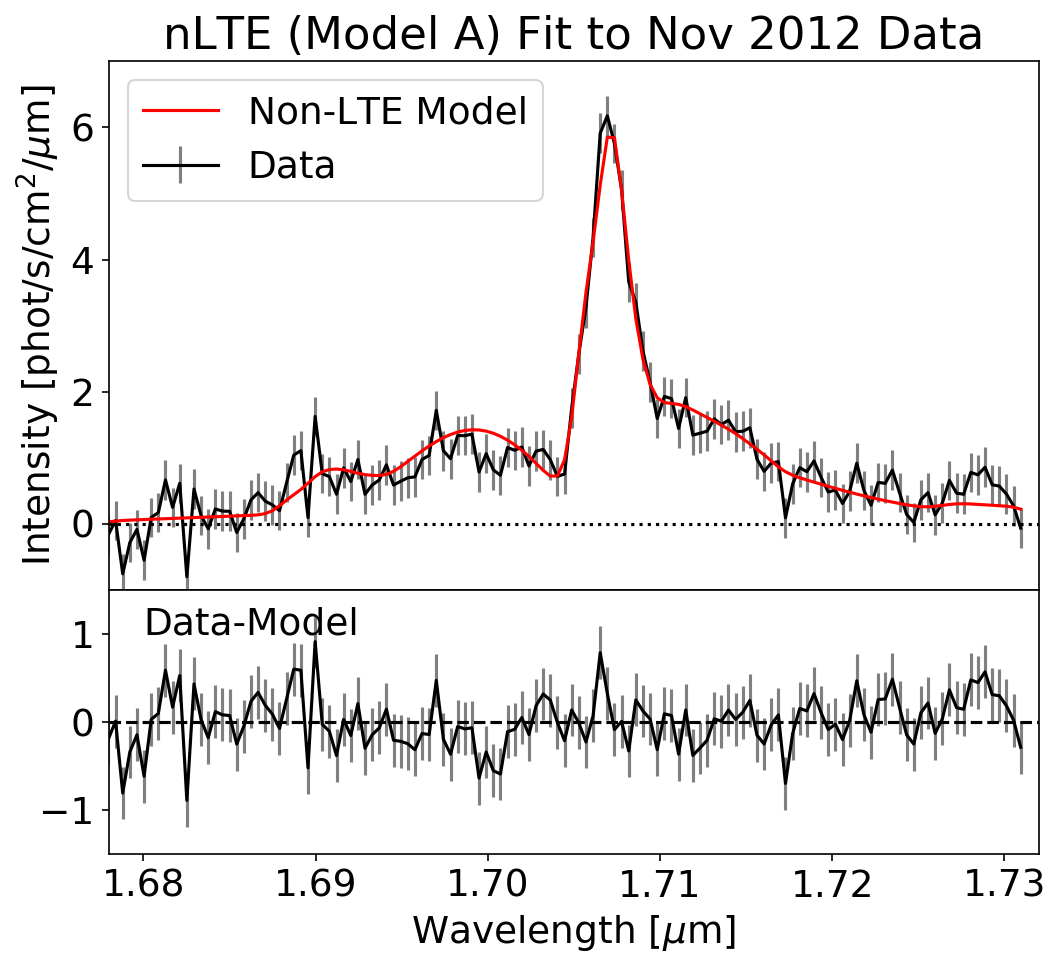}
\includegraphics[width=8cm]{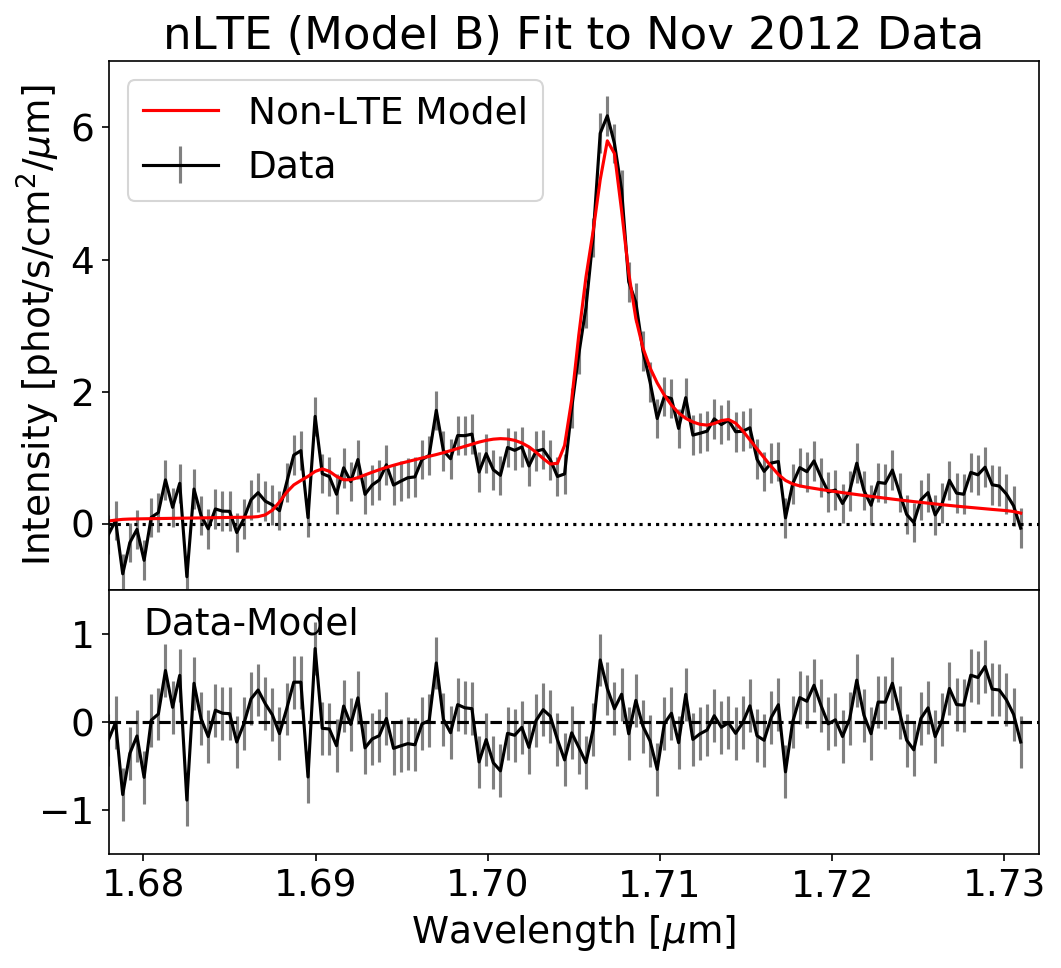}
\caption{Disequilibrium models for the November 2012 dataset that find the lowest $\chi^2$ value for the grid of models tested. (Left) Model that allows high and low rotational states to be populated according to independent temperatures; and (Right) Model with a background atmosphere fixed at the best-fit 2T model parameters, and over-population of high rotational states. \label{fig:SOline_nLTE}}
\end{figure}

For the disequilibrium models that populate high and low rotational states according to different temperatures (nLTE Model A), the $\chi^2$ statistic for the best-fit model for each temperature pair is shown in Figure \ref{fig:x2grid_nLTE}, normalized to the number of datapoints minus the number of free parameters in the fit. The figure demonstrates a preference for a lower temperature around 500 K, although the temperature of the high-T population is poorly constrained. Figure \ref{fig:SOline_nLTE} shows the model with the lowest $\chi^2$ value from this family, as well as the lowest $\chi^2$ model from the fixed background disequilibrium model family. \par
%
%
\subsection{Integrated band strength}
\begin{figure}
\includegraphics[width=18cm]{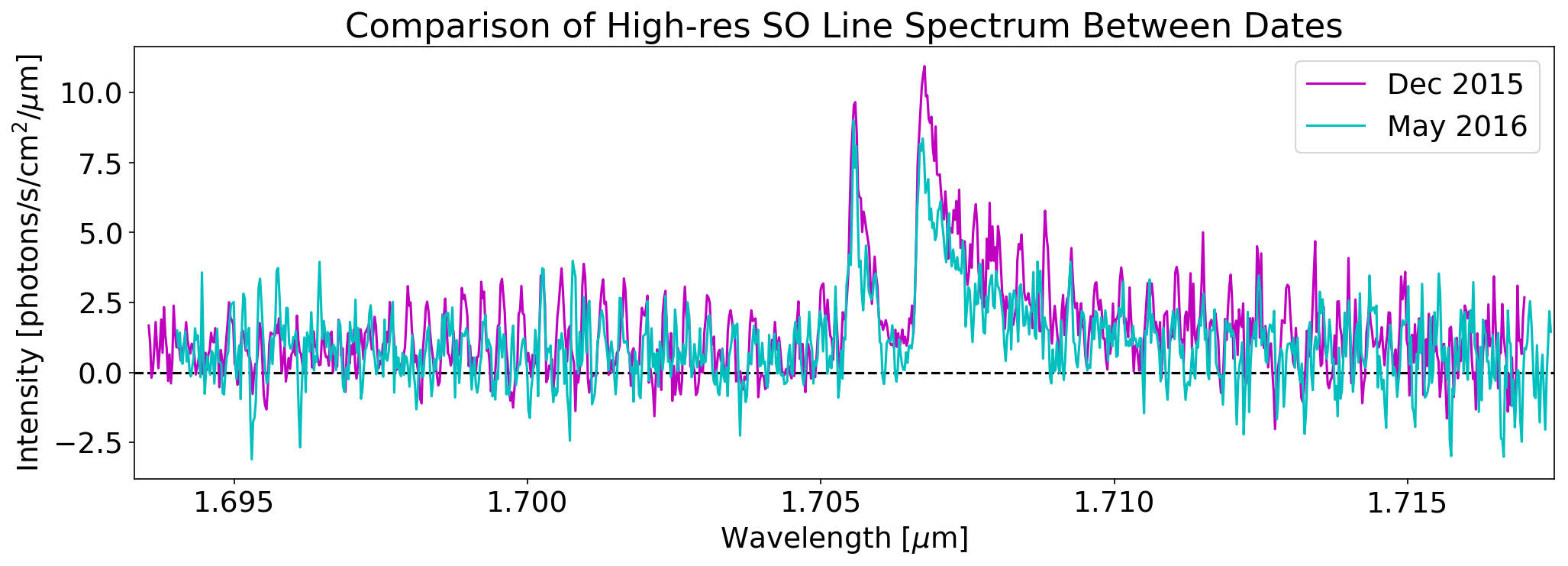}
\caption{Comparison of high-resolution spectra between the two dates of observation, after subtraction of the thermal continuum according to the best-fit parameters for the 2T models (see Table \ref{tbl:results_fits}). Despite the poor weather conditions during the 2016 observations, and the resultant noisy data, the band structure even outside of the main peaks is similar on both dates. \label{fig:spec_highres_comp}}
\end{figure}

Integrating over wavelength across the SO band yields a total photon rate on each date under the assumption of isotropic emission; these values are tabulated in Table \ref{tbl:results} along with the temperature $T_{th}$ and area $A_{th}$ for the preferred thermal continuum models. In November 2012 the emission rate was (2.47$\pm$0.57)$\times$10$^{27}$ photons/s. The emission rates from the high-resolution data, integrated from 1.694-1.717 $\mu$m, are (1.68$\pm$0.39)$\times$10$^{27}$ and (1.16$\pm$0.38)$\times$10$^{27}$ photon/s for Dec 2015 and May 2016 respectively. Note that the numbers from the datasets taken at different spectral resolutions are not directly comparable because the high-resolution datasets cover a smaller portion of the band. By using the November 2012 data, with full spectral coverage of the band, we estimate that the values derived from the high-resolution observations should be increased by $\sim$25\% to yield the integrated emission across the entire band. Table \ref{tbl:results} also gives the breakdown in emission between the two temperature components in the 2T models, and shows that the emission from both components varies considerably between dates. A direct comparison of the high-resolution spectra between 2015 and 2016 is shown in Figure \ref{fig:spec_highres_comp} with thermal continuum removed (as determined from the 2T model fits).  \par
%
\begin{table}
\begin{center}
\caption{Continuum and SO band strength \label{tbl:results}}
\begin{tabular}{c |c c| c c c}
\hline \hline
Date & \multicolumn{2}{c|}{Continuum} & \multicolumn{3}{c}{SO emission$^a$ [10$^{27}$ photons/s]} \\
 & T$_{th}$ [K] & A$_{th}$ [km$^2$] & Total & from T$_1$ & from T$_2$ \\
\hline
Nov 2012 & 870$\pm$6 & 5.5$\pm$1.3 & 2.47$\pm$0.57 & 0.95$\pm$0.22 & 1.52$\pm$0.35 \\
\hline
Dec 2015 & 644$\substack{+40 \\ -2}$ & 240$\substack{+55 \\ -130}$ & 1.68$\pm$0.39 & 0.70$\pm$0.16 & 0.98$\pm$0.23 \\
\hline
May 2016 & 727$\substack{+4 \\ -44}$ & 41$\substack{+50 \\ -14}$ & 1.16$\pm$0.38 & 0.28$\pm$0.09 & 0.88$\pm$0.29 \\
\hline
\end{tabular}
\\
\footnotesize{$^a$Integrated emission for the observed region of the spectrum. The high-resolution data (Dec 2015 and May 2016) do not cover the entire band, and underestimate the total emission by $\sim$25\%. The thermal continuum for these two datasets, and the relative contributions from the two temperature components in all cases, are as derived from the 2T models. The total SO emission is as measured directly from the observed spectra.}
\end{center}
\end{table}

\section{Discussion} \label{sec:disc}
Model fits to the high-spectral-resolution datasets indicate that emission arises from SO gas reservoirs at both high and low temperatures, with high temperatures in the 1000-1500 K range and low temperatures around 75-200 K. For the a$^1\Delta$ excited state of SO, de-excitation due to spontaneous emission is rapid, and the consistent presence of the 1.7-$\mu$m band requires an ongoing production mechanism. De Pater et al. (2002) investigated several excitation mechanisms, and concluded that the most likely scenario involved SO molecules being directly ejected from a volcanic vent in the excited state. This scenario predicts molecules in the excited a$^1\Delta$ state and with high rotational temperatures, matching the high-T component. \par
We now address potential origins for the population of molecules at low rotational temperatures yet in the excited electronic state. The temperatures found for the low-T component are consistent with temperatures previously measured for Io's bulk atmosphere. In Dec 2015 when Io had just entered eclipse, the 172-217 K temperature range is consistent with the 170$\pm$20 K measured for Io's dayside atmosphere at 4 $\mu$m (Lellouch et al. 2015), though somewhat higher than the $<$150 K temperatures inferred from mid-infrared data (Spencer et al. 2005; Tsang et al. 2012). In May 2016, Io had been in eclipse for more than an hour prior to the start of observation, and the gas temperature of 76-119 K is consistent with Io's in-eclipse surface temperature of $\sim$105 K derived from Io's 19-$\mu$m brightness (Tsang et al. 2016). \par
At temperatures of 100-200 K, the thermal energy of the gas ($\sim$0.02 eV) is much smaller than the energy difference between the ground and excited SO gas (0.73 eV), and the cool SO reservoir will not populate the excited state in thermodynamic equilibrium. Molecules with rotational temperatures matching the bulk atmosphere could end up in the a$^1\Delta$ state in two ways: either the electronic state is excited without raising the rotational temperature, or else a fraction of the gas population cools rotationally before de-exciting to the ground electronic state. The former case requires a non-volcanic excitation mechanism, but de Pater et al. (2002) demonstrate that other processes (solar photoexcitation, electron impact excitation of SO or dissociative excitation of SO$_2$, and ionospheric recombination of SO$_2^+$) fail to explain the data. The latter scenario is plausible if the timescale to equilibrate the rotational temperature to the surrounding temperature is on the same order as the timescale of decay from the excited electronic state. \par
We test the consistency of this scenario as follows: we assume that all excited SO gas was initially emitted from a volcanic vent at high rotational temperature, and that the low-temperature emission is sourced from molecules that have equilibrated rotationally to the bulk atmosphere before de-exciting. We then make estimates of the local gas density and pressure from the relative emission from high- and low-T states, and determine if the resultant gas pressures are consistent with past observations. \par
In the scenario described above, the cool excited SO molecules are produced solely through cooling of the hot SO, and are lost solely through spontaneous de-excitation, so that
\begin{equation}
\frac{dn_C}{dt} = -An_C+\frac{n_H}{\tau}
\end{equation}
where $n_H$ and $n_C$ are the number densities of hot and cold excited SO molecules, $A$ is the Einstein coefficient for spontaneous emission from the a$^1 \Delta$ state ($\sim$2.2/sec; Klotz et al. 1984), and $\tau$ is the timescale for collisional cooling. Assuming that the hot and cold SO gas densities remain constant over timescales comparable to the duration of the observations (an assumption supported by the lack of measurable changes in individual spectra over a single observation), the production rate and loss rates of the cool excited SO must be equal:
\begin{equation}
\frac{n_H}{n_C} = A\tau.
\label{eqn:AT}
\end{equation}
\indent The photon rate is equal to the loss rate from spontaneous de-excitation, and the ratio of total photon rate in the 1.7-$\mu$m band to the photon rate from cool SO molecules only is therefore (1+$n_H/n_C$), or (1+A$\tau$) using Equation \ref{eqn:AT}. Using the measured photon rates given in Table \ref{tbl:results}, we derive the timescale for collisional cooling $\tau$. This timescale is the product of the time between collisions $\tau_{coll}$ and the number of collisions to equilibrate ($\sim$tens for exchange of translational and rotational energy). The former is defined as the  ratio of the mean free path $\lambda$ to the mean velocity of molecules in the gas $v_{th}$, while the mean free path is determined by the collisional cross-section $\sigma$ and the density of the gas $n_{bulk}$. In this case, $n_{bulk}$ is the density of the gas in Io's bulk atmosphere, which is the reservoir responsible for the collisional cooling of the SO:
\begin{equation}
\tau_{coll}=\frac{\lambda}{v_{th}}=\frac{1}{n_{bulk}\sigma v_{th}}.
\end{equation}
\indent For the 2015 and 2016 observations respectively, the ratio of the photon rate from all molecules in the a$^1 \Delta$ state to just the low-temperature molecules is 2.4$\substack{+1.4 \\ -0.9}$ and 4.1$\substack{+4.0 \\ -2.0}$, indicating a time to equilibration of around 0.6 and 1.4 seconds, or 0.06 and 0.14 seconds between collisions (assuming $\sim$10 collisions are required to reach rotational equilibrium with the surroundings). These numbers can be translated into number densities using the kinetic velocity of the gas:
\begin{equation}
v_{th}=\sqrt{\frac{8k_BT}{\pi m}},
\end{equation}
where $k_B$ is the Boltzmann constant and $m$ is the molecular mass of SO$_2$, the dominant atmospheric constituent. This yields number densities of 1.3$\substack{+2.5 \\ -0.7}\times$10$^{11}$ and 8.6$\substack{+17 \\ -5.3}\times$10$^{10}$ cm$^{-3}$, or surface pressures around 3.3 and 1.1 nbar, for the 2015 and 2016 observations respectively. Despite the approximate nature of these calculations, these estimates are consistent with past measurements of Io's dayside atmospheric pressure, which have typically been in the few nbar range (Lellouch et al. 1990; 2007), as well as with the density of Io's near-surface dayside atmosphere predicted by numerical models based on observational constraints on column density (Walker et al. 2010). However, these densities are inconsistent with the orders-of-magnitude decrease in both gas density and surface pressure expected for a purely sublimation-supported SO$_2$ atmosphere (Walker et al. 2010). Indeed, recent mid-infrared observations of Io's atmosphere have shown that while the atmospheric SO$_2$ density does drop at night, it only does so by a factor of 5$\pm$2 in density, similarly less than the several orders-of-magnitude decrease predicted by pure vapor pressure equilibrium. \par
During the daytime, Io's scale height of $H_s\sim$10 km is significantly larger than the mean free path (tens of meters). At night, whether or not the atmosphere is collisional, and hence whether the equations above are applicable, depends on the factor by which the gas density decreases. If the density drop is on the order of a factor of five (Tsang et al. 2016), the 5$\times$ longer mean free path is still much smaller than the atmospheric scale height, even accounting for the $\sim$50\% decrease in scale height due to the temperature decrease. \par
The factor of three difference in atmospheric pressure for the post-ingress and pre-egress atmosphere inferred from the 2015 and 2016 observations, respectively, is due largely to the decrease in temperature; the inferred gas density decreases by only 30\%. Note that the lower emission from the low-T component in the May 2016 data (see Table \ref{tbl:results}) can likewise be explained by a lower gas temperature leading to longer time between collisions, and does not require a significant difference in gas density. Although the best-fit densities therefore imply an atmosphere with minimal change ($n_{ingress}/n_{egress}$=1.5), the ratio of densities between ingress and egress is fully consistent with the ratio of 5$\pm$2 measured by Tsang et al. (2016) within uncertainties. Moreover, while Tsang et al. (2016) isolated the density of the SO$_2$ gas, our observations are sensitive to the total bulk atmosphere, which includes other species. The higher $n_{ingress}/n_{egress}$ ratios consistent with our observations correspond to the low end of the acceptable post-ingress temperature range (172 K) and high end of the pre-egress range (119 K), which are also the values most consistent with past work. \par
While the temperature of the low-T gas component in the 2015 and 2016 datasets is therefore consistent with SO gas in rotational equilibrium with the surrounding atmosphere for an atmosphere that cools from $\sim$170 to $\sim$100 K during eclipse, this interpretation is complicated by the fact that fits to individual spectra taken during the eclipse do not show a systematic trend in gas temperature (although signal-to-noise is low in individual spectra). Moreover, even the data taken shortly after Io enters eclipse span a long enough time period that the atmosphere should have cooled if the atmosphere follows the same cooling timescale as the surface. Specifically, the November 2012 data were taken over the first 32 minutes of eclipse, and the December 2015 data during the first 17 minutes of eclipse, while Tsang et al. (2016) find that the change in atmospheric density occurs within the first 20-30 minutes, and that the surface cools from around 130 K to below 110 K within the first 10 minutes of eclipse. If the difference between the low-T gas temperatures derived from the Dec 2015 and May 2016 observations does reflect a cooling of Io's bulk atmosphere, the majority of the cooling must therefore be occurring after Io has been in shadow for $>$20 minutes, implying a delayed cooling of the atmosphere relative to the surface. \par
The 2T models discussed here for the high-resolution datasets are still unable to match the shape of the emission band farther from the line center, as seen in the low-resolution data presented here and in past work (de Pater et al. 2002; Laver et al. 2007; de Pater et al. 2007). In particular, an excess of emission at 1.69 $\mu$m is observed that cannot be fit with equilibrium models. Models that postulate an over-population of high rotational states do a better job at resolving this mismatch. At low densities, such an over-population could arise from either  preferential radiative excitation of the high-J levels or a difference in H{\"o}nl-London factors for high- and low-J lines. The latter would lead to different critical densities for each of the transitions, such that it would be possible for some levels to thermalize while others remain out of equilibrium. We note that at 1.69 $\mu$m emission arises almost exclusively from the $^{\rm S}$R branch of the band, and it is also possible that the over-population is occurring in just this branch, or that the laboratory measurements of oscillator strength are incorrect. As the high-resolution data do not cover these portions of the spectrum and the low-resolution data are unable to discriminate between potential disequilibrium models, a more in-depth characterization of the gas temperature and state in the case of non-LTE is not possible from these datasets. Future observations covering the entire emission band from $\sim$1.67-1.74 $\mu$m at high spectral resolution are required to discriminate between more realistic gas models. \par 
\subsection{Investigating the source of time-variability in total SO emission}
The three new datasets presented here, combined with past work (de Pater et al. 2002; Laver et al. 2007), bring the total number of observations of this emission band to eight, spanning the period from 1999 to 2016. Using all eight observations, we explore possible origins of the emission by determining whether the total amount of gas emission is correlated with (1) Activity at Loki Patera, as was suggested by early observations (de Pater et al. 2002; Laver et al. 2007); (2) Thermally-bright eruptions; (3) Io's mean anomaly, as would be the case if emission were tidally-modulated by the opening and closing of fissures; (4) Io's solar distance, which would be the case if the emission arose from SO in a sublimation-supported bulk atmosphere; (5) The time Io had been in shadow prior to observation, as would be the case if the emission arose from a bulk atmosphere that collapses during eclipse; and (6) Jupiter's System III longitude, which may be the case if electron-impact mechanisms were responsible for the excitation of the SO. In all cases, the values for integrated line strength given for the high-resolution data from 2015 and 2016 are increased by 25\% for comparison with the low-resolution datasets, to correct for the incomplete band coverage.\par
%
For each property, we calculate the Pearson R coefficient to test for correlation, and give both the coefficient and the corresponding p-value. A p-value below 0.05 indicates that the null hypothesis that the SO emission is not correlated with the given property is rejected at the 95\% level. Positive coefficients indicate positive correlations, and negative coefficients indicate anti-correlations.\par
\begin{figure}
\centering
\includegraphics[width=8cm]{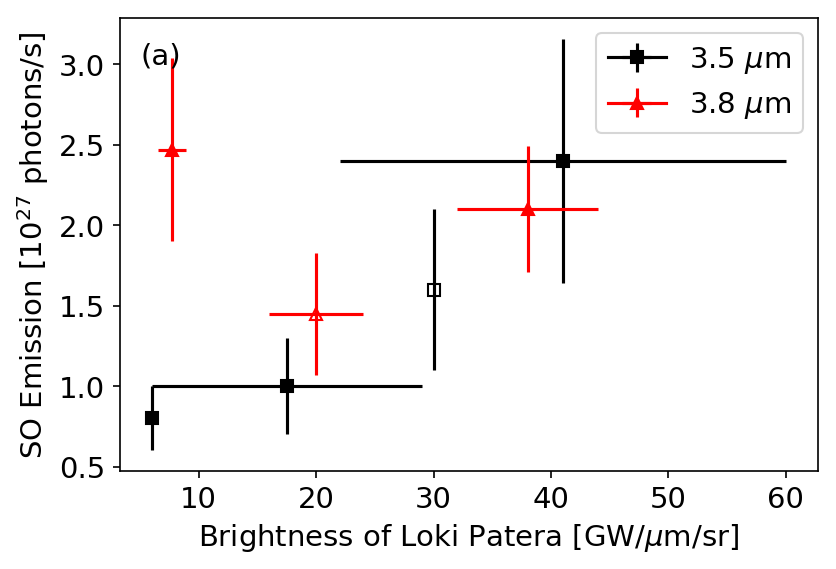}
\includegraphics[width=8cm]{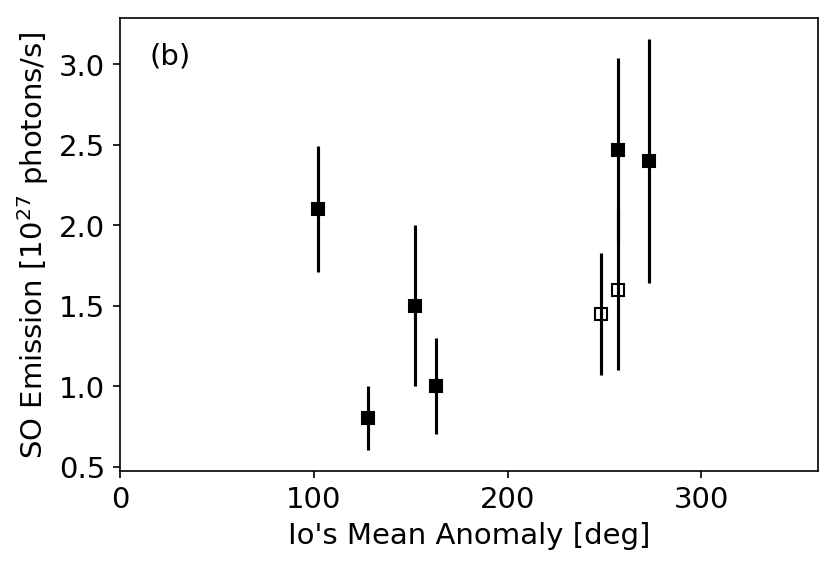}
\includegraphics[width=8cm]{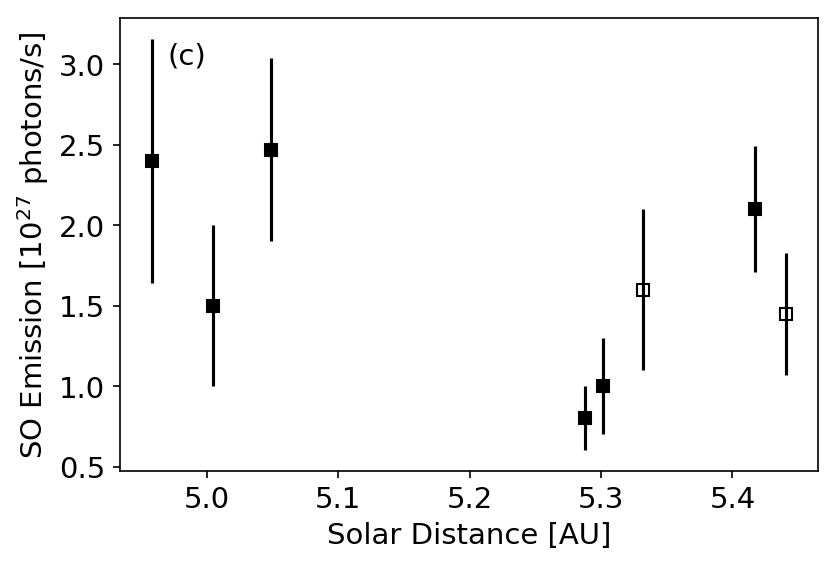}
\includegraphics[width=8cm]{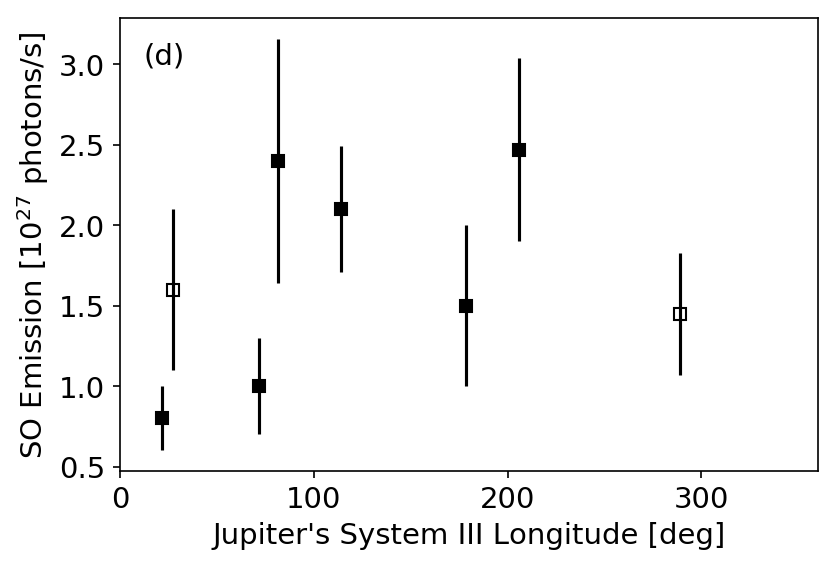}
\caption{Correlation between the total measured SO emission and various properties of Io: (a) Infrared brightness of Loki Patera. Note that the infrared observations were made at slightly different wavelengths, and the brightness at 3.8 $\mu$m is 1-2$\times$ the emission at 3.5 $\mu$m for typical temperatures of 300-500 K; (b) Io's mean anomaly; (c) Io's solar distance; and (d) Io's position in Jupiter's System III longitude. No significant correlation is seen in any plot, indicating that none of these factors individually is the dominant driver behind the total SO emission at 1.7 $\mu$m. Filled symbols correspond to dates when Io was moving into eclipse from sunlight, and open symbols to dates when Io was moving into eclipse from occultation.\label{fig:solardist}}
\end{figure}

\textbf{1. Activity at Loki Patera:}
Initial observations of the 1.7-$\mu$m SO band were suggestive of a correlation between band strength and the level of thermal activity at Loki Patera (de Pater et al. 2002; Laver et al. 2007). Observations that spatially resolved the SO emission did not find a significant contribution from the vicinity of Loki Patera, but the authors note that the observations were made during a quiescent period at this volcano (de Pater et al. 2007). Figure \ref{fig:solardist}a shows the integrated band intensity as a function of Loki Patera's brightness near a wavelength of 3.5 $\mu$m, for all eight observations. While seven of the eight observations follow a suggestive trend, the strong emission seen in November 2012 when Loki Patera was quiescent (and no other volcanoes showed bright thermal activity) deviates significantly from the trend, and there is no statistically-significant correlation (R=0.47; p=0.28). Note that observations at 3.5 and 3.8 $\mu$m are plotted together in Figure \ref{fig:solardist}a, although the flux density at 3.5 $\mu$m may be lower than that at 3.8 $\mu$m by up to 50\% for continuum thermal emission at the cool 300-500 K temperatures typically seen at Loki Patera.\par
\textbf{2. Bright eruptions:}
Thermally-bright eruptions can be accompanied by massive plumes of gas and dust (Geissler and Goldstein, 2007). However, observations of the 1.7-$\mu$m SO band made while a large eruption at Ra Patera was in progress in November 2002 found that only 10-15\% of the total emission was localized to the area around this volcano (de Pater et al. 2007), and the integrated band strength on this date was the weakest of all observations to date. On December 25, 2015 a large, transient eruption was in progress at Amaterasu Patera, and the integrated band strength on this date was also not anomalously large. On the dates with the highest measured SO emission (Sep 24, 1999 and Nov 5, 2012), there was no especially bright activity, although Loki Patera was in the early stages of a bright event during the 1999 observations (de Pater et al. 2017a). We therefore conclude that the thermal emission is not a good indicator of the amount of excited SO emission, and if the SO is of volcanic origin it is not dominated by events with large thermal signatures.\par
\textbf{3. Io's mean anomaly:}
Since the amount of 1.7-$\mu$m SO emission is not dominated by bright volcanic eruptions, perhaps it is driven by volcanic gas emission from fissures that don't have active lava flow components. In such a case, the amount of emission could conceivably be modulated by tidal stresses opening and closing cracks such as in the case of Enceladus (Hedman et al. 2013; Hurford et al. 2007). We test this hypothesis by looking at integrated band strength as a function of Io's mean anomaly; Figure \ref{fig:solardist}b illustrates that such a correlation is not present at a detectable level (R=0.47; p=0.24).\par
\textbf{4. Io's solar distance:}
The density of Io's bulk SO$_2$ atmosphere is correlated with solar distance, which suggests that there is a significant component that is controlled by sublimation (Tsang et al. 2012). If the 1.7-$\mu$m SO emission arises from SO molecules in Io's bulk atmosphere, which is primarily sourced from SO$_2$, we might expect to see a similar correlation. Figure \ref{fig:solardist}c shows the integrated SO band intensity as a function of Jupiter's solar distance; although there is a weak trend in the expected direction, a best-fit line to the points is consistent with zero slope to within 1$\sigma$ and there is no statistically-significant correlation (R=-0.46; p=0.25).\par
%
\textbf{5. Time since in sunlight:}
Of the eight observations, all but two were taken shortly after Io had moved from sunlight into eclipse. For the two datasets taken with Io moving from Jupiter occultation into eclipse (i.e. having been in shadow for $\geq$1 hour prior to observation), the integrated line strengths were 1.6$\pm$0.5 and 1.45$\pm$0.38 $\times$10$^{27}$ photons/sec (datasets from 2003 and 2016 respectively), which are both average line strengths relative to the other datasets. It therefore appears that the time that Io has been in shadow prior to observation is not the dominant control on the intensity of emission and hence that, if the bulk SO atmosphere is sourced photolytically and is lost to surface interactions over the timescale of an eclipse, then the observed SO emission must be primarily of volcanic origin. \par
\textbf{6. Jupiter's System III longitude:}
If SO molecules were excited to the a$^1\Delta$ state via electron impact processes (a possibility considered and rejected by de Pater et al. 2002), we might expect to see a correlation in emission with Jupiter's System III longitude. The SO band strength is plotted against Io's position in Jupiter's System III Longitude in Figure \ref{fig:solardist}d, illustrating that such a correlation is not observed (R=0.29; p=0.49). \par
\par
\section{Conclusions}\label{sec:conc}
We observed Io in Jupiter eclipse on three occasions in 2012-2016 and detected the 1.7-$\mu$m rovibronic band associated with the forbidden a$^1\Delta \rightarrow {\rm X}^3 \Sigma^-$ transition of SO. On two of these dates, observations were made at a spectral resolution of R$\sim$15,000, an order of magnitude improvement over all prior datasets. The high spectral resolution of these data permits a more detailed modeling treatment than has been possible in the past. Analysis of the spectra indicates a contribution from gas components at both high and low rotational temperatures. The high-temperature component is consistent with volcanic gas emission (1000-1500 K), while the low-temperature component is consistent with the temperature of Io's bulk atmosphere: 172-217 K during observations of Io just entering eclipse (in December 2015), and 76-119 K after Io had been in shadow for an hour (in May 2016). This is suggestive of a hot volcanic gas that has partially equilibrated rotationally to the bulk atmospheric temperature prior to de-exciting to the ground electronic state, a scenario which is consistent with the radiative properties of SO for a gas density of $\sim 10^{11}$ cm$^{-3}$ and a pressure of $\sim$1-3 nbar in the emitting region, if the atmosphere cools through the eclipse. These values are consistent with past observational and modeling work for Io's dayside temperature (e.g. Walker et al. 2010; Lellouch et al. 2007). \par
This interpretation is challenged by the fact that Io's surface cools after entering eclipse over a timescale of $\sim$10 minutes (Tsang et al. 2016), while individual spectra taken throughout the first 30 minutes of eclipse do not show any systematic trends in gas temperature (though we note that if such a change were subtle, it would not be distinguished in the individual spectra due to low signal-to-noise). If the low-temperature SO emission is indeed probing the bulk atmospheric temperature, the atmospheric cooling must be delayed relative to that of the surface. In addition, the overall band strengths and best-fit gas densities are very similar between the two observing dates. This would seem to contradict recent evidence for the collapse of Io's bulk SO$_2$ atmosphere in eclipse (Tsang et al. 2016). However, we note that uncertainties on the inferred densities are high, and rely on several assumptions; a collapse by a factor of 5$\pm$2, as found by Tsang et al. (2016), is fully consistent with our results.\par
These three datasets bring the total number of detections of this emission band to eight. Analysis of the total band strength across all eight dates does not find any significant correlation between the observed SO emission and incident sunlight, Io's orbital phase, time since Io was last in sunlight, Jupiter's System III longitude, nor thermal hot spot activity. If the SO producing the emission band is indeed of volcanic origin, this indicates that thermal hot spot activity is not a good indicator of the gas emission. \par
The two-temperature models that provide a good fit to the high-resolution data are unable to reproduce an excess of emission near 1.69 $\mu$m that is seen in the low-resolution data, which cover a wider spectral range. Simple non-LTE models are able to match this emission by over-populating high rotational states, but a detailed analysis is limited by the coarse spectral resolution. Future observations with high spectral resolution across the entire 1.67-1.74 $\mu$m region will allow for a more in-depth characterization of the thermodynamic equilibrium state of the gas, which reflects on the gas origins and potentially the vent conditions. Continued observations with high spectral resolution after Io has been in shadow for differing time intervals would confirm or refute the tentative correlation between the temperature of the low-T gas component and the amount of time Io has been in shadow; if the correlation is confirmed, such studies would then provide a new way of studying how Io's atmospheric temperature and density drop at night. \par
\centerline {\bf Acknowledgments}
K. de Kleer is supported by the Heising-Simons Foundation \textit{51 Pegasi b} postdoctoral fellowship; this work was also partially supported by the National Science Foundation grant AST-1313485 to UC Berkeley. This work made use of the JPL Solar System Dynamics high-precision ephemerides through the HORIZONS system. Data were obtained with the W.M. Keck Observatory, which is operated by the California Institute of Technology, the University of California, and the National Aeronautics and Space Administration. The Observatory was made possible by the generous financial support of the W.M. Keck Foundation. Data were also obtained at the Gemini Observatory, which is operated by the Association of Universities for Research in Astronomy, Inc., under a cooperative agreement with the NSF on behalf of the Gemini partnership: the National Science Foundation (United States), the National Research Council (Canada), CONICYT (Chile), Ministerio de Ciencia, Tecnolog\'{i}a e Innovaci\'{o}n Productiva (Argentina), and Minist\'{e}rio da Ci\^{e}ncia, Tecnologia e Inova\c{c}\~{a}o (Brazil). The authors extend special thanks to those of Hawaiian ancestry on whose sacred mountain we are privileged to be guests. Without their generous hospitality, none of the observations presented would have been possible. \par
\appendix
\section{Statistical methods for parameter and uncertainty estimation} \label{sec:appendix}
\begin{figure}[ht]
\includegraphics[width=14cm]{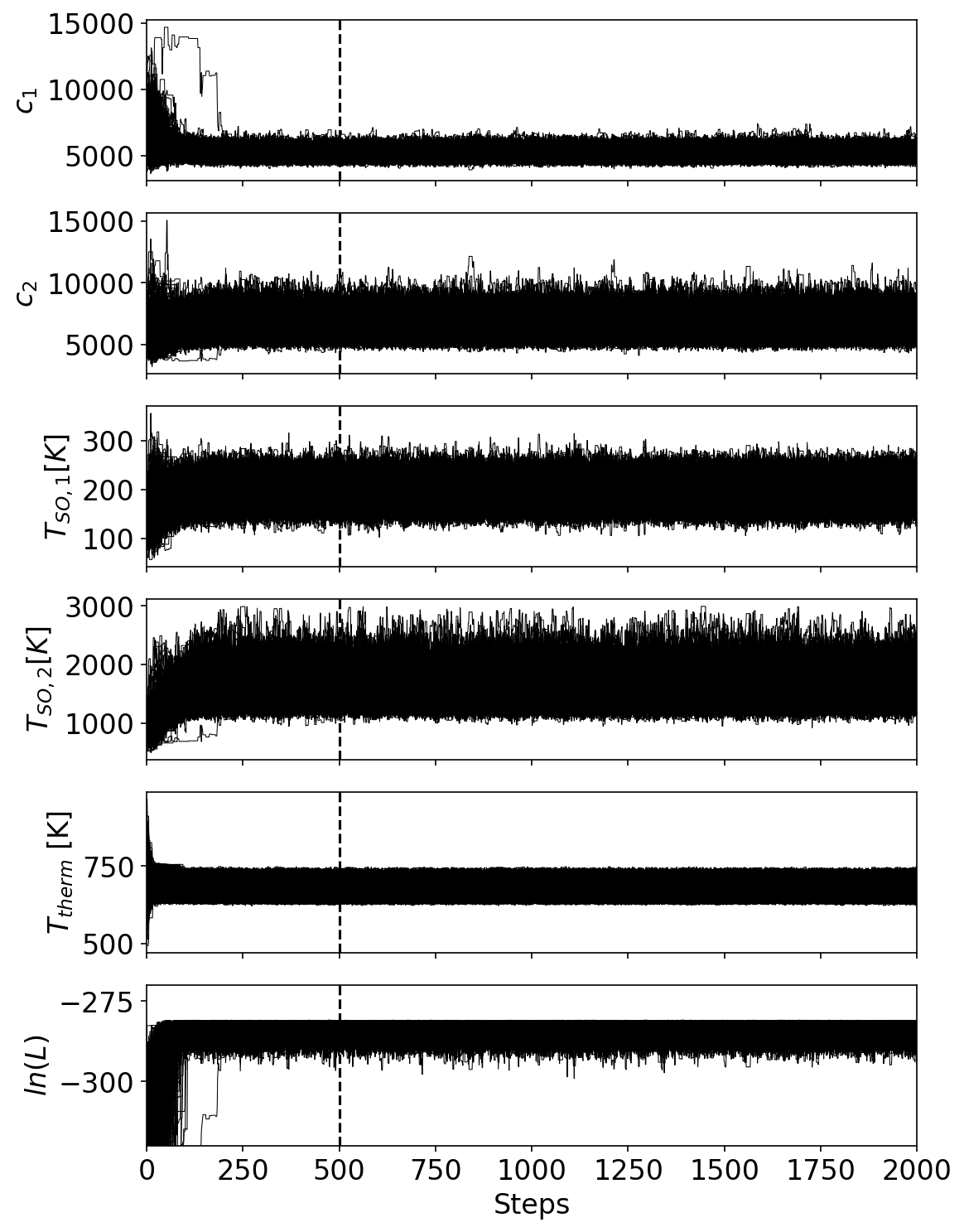}
\caption{MCMC simulations: plots show the parameter values taken on by each of 1000 simultaneous chains at each of 2000 steps. Only points to the right of the dashed vertical line are used in constructing the posterior distributions. \label{fig:walkers}}
\end{figure}

\begin{figure}[ht]
\includegraphics[width=14cm]{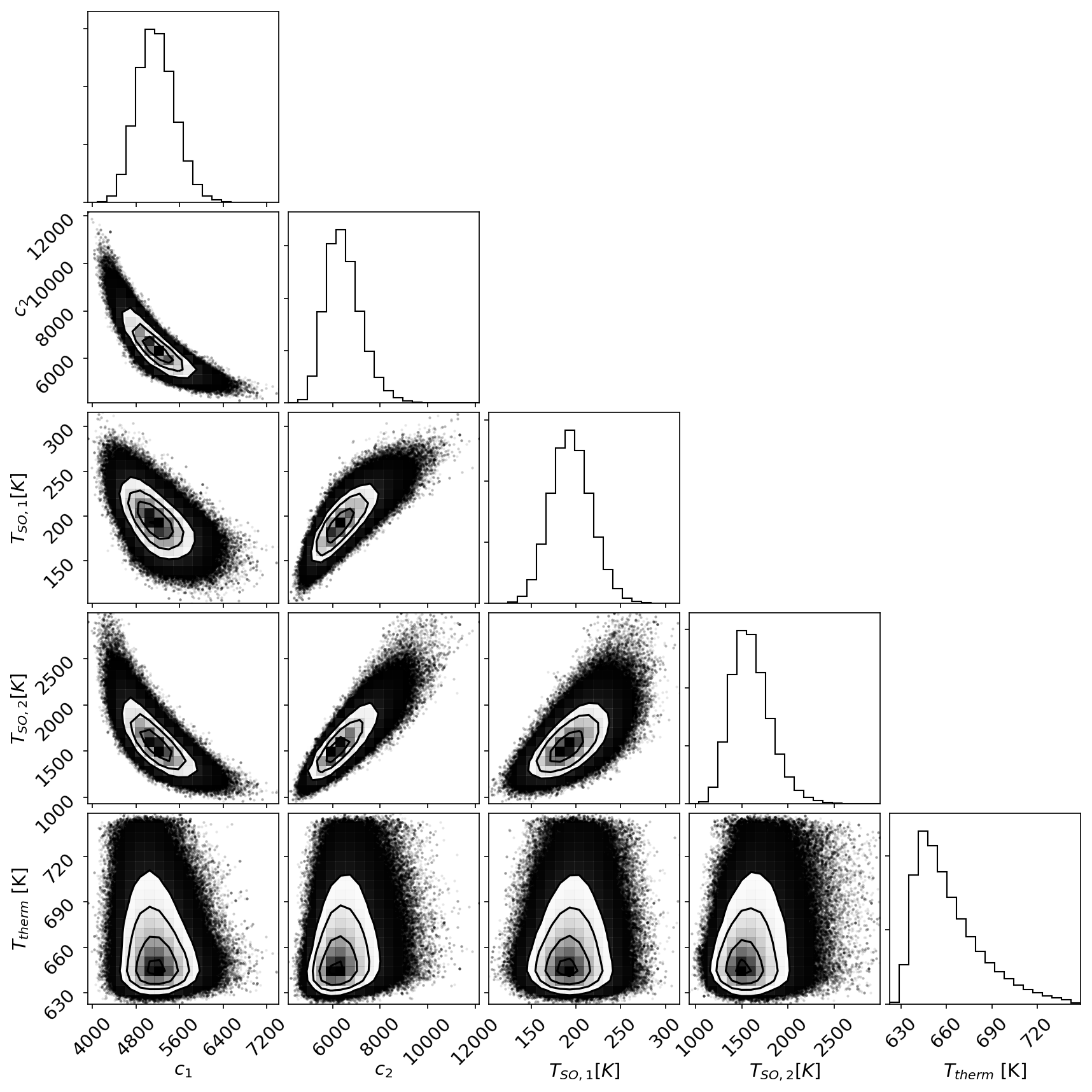}
\caption{Single and joint probability distributions for all parameters in the 2T fits to the December 2015 data. \label{fig:corner}}
\end{figure}

The best-fit parameter values for each of the equilibrium models (1T and 2T) described in the main text are determined through a likelihood maximization routine, and are presented as the maximum likelihood values in Table \ref{tbl:results_fits}. The uncertainties are estimated using an affine-invariant ensemble Markov Chain Monte Carlo (MCMC) simulation, via the Python \textbf{emcee} package (Goodman \& Weare 2010; Foreman-Mackey et al. 2012). Such simulations explore parameter space to sample the underlying probability distribution for each parameter. An example is shown for the 2T model fit to the December 2015 data in Figures \ref{fig:walkers} and \ref{fig:corner}. Figure \ref{fig:walkers} shows the value of each parameter as a function of step, for 1000 simultaneous chains; only values to the right of the dashed vertical line are used to generate the probability distributions, ensuring that distributions use only post-convergence values. Figure \ref{fig:corner} shows the joint and single probability distributions for each parameter. Because these simulations recover the probability distribution for each parameter under the modeling assumptions, the expectation of each parameter and its 1-$\sigma$ uncertainty can be read from the distribution directly. However, in the case of asymmetric probability distributions, the 50th quantile may deviate from the maximum likelihood value. The 50th quantile is therefore not quoted, but the 16th and 84th quantile are given in Table \ref{tbl:results_fits} as the credible interval for each parameter. \\
\subsection{Modeling the thermal continuum}
On each date, the continuum level in the vicinity of 1.7 $\mu$m is set by thermal emission from high-temperature volcanism, and its intensity and spectral shape vary between observations according to Io's volcanic activity. The low-resolution data from 2012 contain sufficient spectral coverage outside of the SO band to fit for the thermal background and subtract it from the data prior to the gas emission retrievals (see Figure \ref{fig:spec_2012}). This is not the case for the high-resolution datasets from 2015 and 2016; for the data from these dates, the thermal continuum is fit simultaneously with the gas properties. \par
In all cases, the thermal background is modeled as a single emitting area at a single temperature; these models fit the data well, justifying the use of such a simplified formulation. The initial fits to the high-resolution data treat the emitting temperature and area ($T_{th}$ and $A_{th}$) as independent free parameters in the fits. However, on both dates the simulations find these parameters to be completely degenerate, indicating that for the best-fitting temperature regime, the minor changes in spectral shape between neighboring temperatures do not significantly affect the spectrum. The emitting area as a function of temperature is well-described by a fourth degree polynomial; after determining the polynomial coefficients for each dataset and model, the emitting area is directly calculated from the temperature in the fits, reducing the number of free parameters by one. 
%
%
%
\begin{figure}[ht]
\includegraphics[width=10cm]{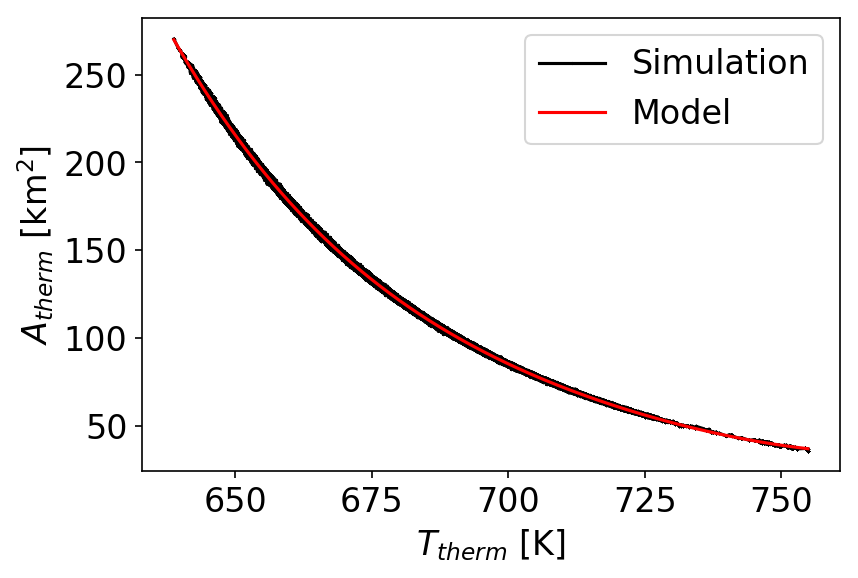}
\caption{The joint probability distribution of the thermal continuum temperature and area from MCMC simulations using the 2T model and the Dec 2015 data. The parameters are found to be degenerate; the modeled $A_{th}$($T_{th}$) is shown in red; this curve is used to calculate the emitting area directly from the temperature, reducing the number of free parameters in the fits. \label{fig:therm_param_ex}}
\end{figure}
\clearpage
References:\\
\begin{itemize}
\item[]{Andrae, R., Schulze-Hartung, T., Melchior, P. 2010. Dos and don'ts of reduced chi-squared. arXiv:1012.3754.}
\item[]{Cantrall, C., de Kleer, K., de Pater, I. et al. 2018. Variability and geologic associations of volcanic activity on Io in 2001-2016. Icarus, 312, 267-294.}
\item[]{Castelli, F., Kurucz, R.L., 2004. New grids of ATLAS9 model atmospheres. astro-ph/0405087.}
\item[]{de Kleer, K., de Pater, I., 2016. Time-variability of Io's volcanic activity from near-IR adaptive optics observations on 100 nights in 2013-2015. Icarus 280, 378-404.}
\item[]{de Pater, I., Roe, H.G., Graham, J.R., Strobel, D.F., Bernath, P., 2002. Detection of the forbidden SO $a^1\Delta \rightarrow X^3 \Sigma ^-$ rovibronic transition on Io at 1.7 $\mu$m. Icarus 156, 296-301.}
\item[]{de Pater, I., Laver, C., Marchis, F., Roe, H., Macintosh, B.A. Spatially resolved observations of the forbidden SO $a^1\Delta \rightarrow X^3 \Sigma ^-$ rovibronic transition on Io during an eclipse and a volcanic eruption at Ra Patera. Icarus 191, 172-182 (2007).}
\item[]{de Pater, I., de Kleer, K., Davies, A.G., \'Ad\'amkovics, M., 2017a. Three decades of Loki Patera observations. Icarus 297, 265-281.}
\item[]{de Pater, I., de Kleer, K., \'Ad\'amkovics, M., 2017b. Spatial distribution of the forbidden 1.707 $\mu$m rovibronic emission on Io. AAS DPS Meeting \#49, ID 214.01.}
\item[]{de Pater, I., de Kleer, K., \'Ad\'amkovics, M. Spatial distribution of the forbidden 1.707 $\mu$m rovibronic emission on Io. \textit{in prep.}}
\item[]{Feaga, L.M., McGrath, M., Feldman, P.D., 2009. Io's dayside SO$_2$ atmosphere. Icarus 201, 570-584.}
\item[]{Geissler, P.E., Goldstein, D.B., 2007. Plumes and their deposits. In: \textit{Io After Galileo}, ed. Lopes, R.M., Spencer, J.R. p. 163-192.}
\item[]{Hedman, M.M., Gosmeyer, C.M., Nicholson, P.D., et al. 2013. An observed correlation between plume activity and tidal stresses on Enceladus. Nature 500, 182-184.}
\item[]{Hurford, T.A., Helfenstein, P., Hoppa, G.V., Greenberg, R., Bills, B.G., 2007. Eruptions arising from tidally controlled periodic openings of rifts on Enceladus. Nature 447, 292-294.}
\item[]{Jessup, K.L., Spencer, J.R., Ballester, G.E., et al. 2004. The atmospheric signature of Io's Prometheus plume and anti-Jovian hemisphere: Evidence for sublimation atmosphere. Icarus 169, 197-215.}
\item[]{Jessup, K.L., Spencer, J.R., 2015. Spatially resolved HST-STIS observations of Io's dayside equatorial atmosphere. Icarus 248, 165-189.}
\item[]{Klotz, R., Marian, C.M., Peyerimhoff, S.D., Hess, N.A., Buenker, R.J. 1984. Calculation of spin-forbidden radiative transitions using correlated wavefunctions" Lifetimes of $b^1\Sigma^+$, $a^1\Delta$ states in O$_2$, S$_2$, and SO. Chem Phys. 89, 223-236.}
\item[]{Laver, C., de Pater, I., Roe, H.G., Strobel, D.F., 2007. Temporal behavior of the SO 1.707 $\mu$m ro-vibronic emission band in Io's atmosphere. Icarus 189, 401-408.}
\item[]{Larkin, J., Barczys, M., Krabbe, A., et al. 2006. OSIRIS: a diffraction limited integral field spectrograph for Keck. Proc. SPIE 6269, 1-5.}
\item[]{Lellouch, E., Belton, M., de Pater, I., Gulkis, S., Encrenaz, T. Io's atmosphere from microwave detection SO2. Nature 346, 639-641 (1990).}
\item[]{Lellouch, E., McGrath, M.A., Jessup, K.L., 2007. Io's atmosphere. In: \textit{Io After Galileo}, ed. Lopes, R.M., Spencer, J.R. p. 231-264.}
\item[]{Lellouch, E., Ali-Dib, M., Jessup, K.-L. et al. 2015. Detection and characterization of Io's atmosphere from high-resolution 4-$\mu$m spectroscopy. Icarus 253, 99-114.}
\item[]{Lim, P.L., Diaz, R.I., Laidler, V., 2015. PySynphot User's Guide (Baltimore, MD: STSci), http://pysynphot.readthedocs.io/en/latest/.}
\item[]{McLean, I.S., Becklin, E.E., Bendiksen, O., et al. 1998. Design and development of NIRSPEC: a near-infrared echelle spectrograph for the Keck II telescope. Proc. SPIE 3354, 566-578.}
\item[]{Mieda, E., Wright, S.A., Larkin, J.E., 2014. Efficiency measurements and installation of a new grating for the OSIRIS spectrograph at Keck Observatory. PASP 126, 250-263.}
\item[]{Moullet, A., Lellouch, E., Moreno, R. et al., 2013. Exploring Io's atmospheric composition of APEX: First measurement of $^{34}$SO$_2$ and tentative detection of KCl. ApJ 776, 32-40.}
\item[]{Retherford, K.D., Spencer, J.R., Stern, S.A. et al., 2007. Io's atmospheric response to eclipse: UV aurorae observations. Science 318, 237-241.}
\item[]{Roth, L., Saur, J., Retherford, K.D., Strobel, D.F., Spencer, J.R. 2011. Simulation of Io's auroral emission: Constraints on the atmosphere in eclipse. Icarus 214, 495-509.}
\item[]{Schneider, N.M., Bagenal, F. 2007. Io's neutral clouds, plasma torus, and magnetospheric interaction. In: \textit{Io After Galileo}, ed. Lopes, R.M., Spencer, J.R. p. 265-286.}
\item[]{Setzer, K.D., Fink, E.H., Ramsay, D.A. 1999. High-resolution Fourier-transform study of the $b^1\Sigma^+ \rightarrow X^3\Sigma^-$ and $a^1\Delta \rightarrow X^3\Sigma^-$ transitions of SO. Jour. Mol. Spec. 198, 163-174.}
\item[]{Spencer, J.R., Lellouch, E., Richter, M.J. et al. 2005. Mid-infrared detection of large longitudinal asymmetries in Io's SO$_2$ atmosphere. Icarus 75, 207-232.}
\item[]{Tsang, C.C.C., Spencer, J.R., Lellouch, E., et al. 2012. Io's atmosphere: Constraints on sublimation support from density variations on seasonal timescales using NASA IRTF/TEXES observations from 2001 to 2010. Icarus 217, 277-296.}
\item[]{Tsang, C.C.C., Spencer, J.R., Lellouch, E., Lopes-Valverde, M.A., Richter, J.J. 2016. The collapse of Io's primary atmosphere in Jupiter eclipse. JGR: Planets 121, 1400-1410.}
\end{itemize}
\end{document}